\pdfoutput=1
\documentclass[aps,prd,amsmath,floats,floatfix,twocolumn,superscriptaddress,nofootinbib,showpacs,longbibliography]{revtex4-1}

\usepackage[T1]{fontenc}
\usepackage[utf8]{inputenc}
\usepackage{lmodern}
\usepackage{verbatim}

\usepackage[dvipsnames, usenames]{xcolor}
\definecolor{linkcolor}{rgb}{0.0,0.3,0.5}
\usepackage[hypertexnames=false, unicode, colorlinks=true, linkcolor=linkcolor,
citecolor=linkcolor, filecolor=linkcolor,urlcolor=linkcolor,
pdfusetitle]{hyperref}

\usepackage[all]{hypcap}
\usepackage{graphicx}
\usepackage{xspace}
\usepackage{amssymb}
\usepackage[normalem]{ulem} %for \sout
\usepackage{bm} % boldmath
\usepackage{microtype}

\usepackage{enumitem}

\usepackage[english]{babel}
\graphicspath{%
  {figs/}%
}

\DeclareMathAlphabet{\mathpzc}{OT1}{pzc}{m}{it}

\newcommand{\roughly}{\mathchar"5218\relax\,} % Different from \sim in spacing
\newcommand{\into}{\!\times\!\relax} % Different from \times in spacing

\newcommand{\h}{\mathpzc{h}}
\newcommand{\hlm}{\mathpzc{h}_{\ell m}}

\newcommand{\htwotwo}{\mathpzc{h}_{22}}

\newcommand{\dd}{\mathrm{d}}

\newcommand{\etal}{{\itshape et al.}}

\newcommand{\eOrb}{e_{\Omega_{\text{orb}}}}
\newcommand{\egw}{e_{\text{gw}}}
\newcommand{\etwotwo}{e_{\omega_{22}}}
\newcommand{\eNewtonian}{e_{\text{Newt}}}
\newcommand{\lgw}{l_{\text{gw}}}
\newcommand{\omegaA}{\omega^{\text{a}}_{22}}
\newcommand{\omegaP}{\omega^{\text{p}}_{22}}

\newcommand{\fLow}{f_{\text{low}}}

\newcommand{\tLow}{t_{\text{low}}}
\newcommand{\omegatwotwo}{\omega_{22}}
\newcommand{\omegaOrb}{\Omega_{\text{orb}}}
\newcommand{\omegaOrbA}{\Omega^{\text{a}}_{\text{orb}}}
\newcommand{\omegaOrbP}{\Omega^{\text{p}}_{\text{orb}}}
\newcommand{\avgOmega}{\langle\omega_{22}\rangle}
\newcommand{\omegaMean}{\omega^{\text{mean}}_{22}}

\newcommand{\omegaFitP}{\omega^{\text{fit},\text{p}}_{22}}
\newcommand{\omegaFitA}{\omega^{\text{fit},\text{a}}_{22}}

\newcommand{\avgTP}{\langle t\rangle^{\text{p}}}
\newcommand{\avgTA}{\langle t\rangle^{\text{a}}}
\newcommand{\phitwotwo}{\phi_{22}}
\newcommand{\eEOB}{e_{\text{eob}}}

\newcommand{\egeo}{e_{\text{geo}}}
\newcommand{\zeroA}{A_{22}^{\text{circ}}}
\newcommand{\zeroOmega}{\omega_{22}^{\text{circ}}}
\newcommand{\tref}{t_{\text{ref}}}
\newcommand{\fref}{f_{\text{ref}}}

\newcommand{\mAmp}{\texttt{Amplitude}}
\newcommand{\mFreq}{\texttt{Frequency}}
\newcommand{\mResAmp}{\texttt{ResidualAmplitude}}
\newcommand{\mResFreq}{\texttt{ResidualFrequency}}
\newcommand{\mFreqFits}{\texttt{FrequencyFits}}
\newcommand{\mAmpFits}{\texttt{AmplitudeFits}}
\newcommand{\resAmp}{\Delta A_{22}}
\newcommand{\resOmega}{\Delta \omega_{22}}
\newcommand{\Atwotwo}{A_{22}}
\newcommand{\tStart}{t_{0}}
\newcommand{\tStartHat}{\widehat{t}_{0}}

\newcommand{\SEOB}{\texttt{SEOBNRv4EHM}}

\newcommand{\SEOBNRE}{\texttt{SEOBNRE}}
\newcommand{\TEOB}{\texttt{TEOBResumS-DALI}}

\newcommand{\EccentricTD}{\texttt{EccentricTD}}
\newcommand{\PhenomT}{\texttt{IMRPhenomT}}

\newcommand{\SpEC}{\texttt{SpEC}}
\newcommand{\python}{\texttt{Python}}
\newcommand{\package}{\texttt{gw\_eccentricity}~\cite{gw_eccentricity}}

\newcommand{\tP}{t^{\text{p}}}
\newcommand{\tA}{t^{\text{a}}}
\newcommand{\dm}{(2, 2)}

\newcommand{\hatT}{\hat{t}}
\newcommand{\eccfitTp}{T}
\newcommand{\eccfitT}{t_{\rm merg}}
\newcommand{\eccfitTmid}{t_{\rm mid}}

\newcommand{\uu}{U}

\newcommand{\cPeri}{blue}
\newcommand{\cApo}{pink}

\newcommand{\cVline}{gray}

\newcommand{\markApo}{square}
\newcommand{\markPeri}{circle}

\begin{document}

\title{Defining eccentricity for gravitational wave astronomy}

\newcommand{\Cornell}{\affiliation{Cornell Center for Astrophysics
    and Planetary Science, Cornell University, Ithaca, New York 14853, USA}}
\newcommand\CornellPhys{\affiliation{Department of Physics, Cornell
    University, Ithaca, New York 14853, USA}}
\newcommand\Caltech{\affiliation{TAPIR 350-17, California Institute of
    Technology, 1200 E California Boulevard, Pasadena, CA 91125, USA}}
\newcommand{\AEI}{\affiliation{Max Planck Institute for Gravitational Physics
    (Albert Einstein Institute), D-14476 Potsdam, Germany}}
\newcommand{\UMassD}{\affiliation{Department of Mathematics,
    Center for Scientific Computing and Data Science Research,
    University of Massachusetts, Dartmouth, MA 02747, USA}}
\newcommand\Olemiss{\affiliation{Department of Physics and Astronomy,
    The University of Mississippi, University, MS 38677, USA}}
\newcommand{\Bham}{\affiliation{School of Physics and Astronomy and Institute
    for Gravitational Wave Astronomy, University of Birmingham, Birmingham, B15
    2TT, UK}}
\newcommand{\ICTS}{\affiliation{International Centre for Theoretical Sciences,
    Tata Institute of Fundamental Research, Bangalore 560089, India}}
\newcommand{\SNU}{\affiliation{Department of Physics and Astronomy,
    Seoul National University, Seoul 08826, Korea}}
\newcommand{\NBIA}{
	\affiliation{Niels Bohr International Academy, Niels Bohr Institute, Blegdamsvej 17, 2100 Copenhagen, Denmark}}

\author{Md Arif Shaikh}
\email{arifshaikh.astro@gmail.com}
\SNU
\ICTS

\author{Vijay Varma}
\email{vijay.varma@aei.mpg.de}
\thanks{Marie Curie Fellow}
\AEI
\UMassD

\author{Harald P. Pfeiffer}
\AEI

\author{Antoni Ramos-Buades}
\AEI

\author{Maarten van de Meent}
\AEI
\NBIA

% Because hyperref only gets the *last* author, we need to be explicit.
\hypersetup{pdfauthor={Shaikh et al.}}

\date{\today}

%==========================================================================
\begin{abstract} Eccentric compact binary mergers are significant scientific
targets for current and future gravitational wave observatories. To detect and
analyze eccentric signals, there is an increasing effort to develop waveform
models, numerical relativity simulations, and parameter estimation frameworks
for eccentric binaries. Unfortunately, current models and simulations use
different internal parameterisations of eccentricity in the absence of a unique
natural definition of eccentricity in general relativity, which can result in
incompatible eccentricity measurements. In this paper, we adopt a
standardized definition of eccentricity and mean anomaly based solely on
waveform quantities, and make our implementation publicly available
through an easy-to-use \texttt{Python} package, \texttt{gw\_eccentricity}.
This definition is free of gauge ambiguities, has the correct Newtonian limit,
and can be applied as a postprocessing step when comparing eccentricity
measurements from different models. This standardization puts all models and
simulations on the same footing and enables direct comparisons between
eccentricity estimates from gravitational wave observations and astrophysical
predictions. We demonstrate the applicability of this definition and the
robustness of our implementation for waveforms of different origins, including
post-Newtonian theory, effective one body, extreme mass ratio inspirals, and
numerical relativity simulations.  We focus on binaries without spin-precession
in this work, but possible generalizations to spin-precessing binaries are
discussed.
\end{abstract}

\maketitle

%==========================================================================
\section{Introduction}
\label{sec:introduction}
The gravitational wave (GW) detectors
LIGO~\cite{TheLIGOScientific:2014jea} and
Virgo~\cite{TheVirgo:2014hva} have observed a total of $\roughly 90$
compact binary coalescences so far~\cite{LIGOScientific:2021djp},
which includes binary black holes (BHs)~\cite{Abbott:2016blz}, binary
neutron stars (NSs)~\cite{TheLIGOScientific:2017qsa} and BH-NS
binaries~\cite{LIGOScientific:2021qlt}. One of the key goals of GW
astronomy is to understand how such compact binaries form in
nature. The astrophysical source properties inferred from the GW
signals carry valuable clues about the origin of these binaries. In
particular, the spins of the compact objects and the eccentricity of
the orbit are powerful GW observables for this purpose.

If the spins are aligned with the orbital angular momentum, the
orbital plane remains fixed throughout the evolution. If the spins are
tilted on the other hand, the spins interact with the orbit, causing
the orbital plane to precess on a timescale of several
orbits~\cite{Apostolatos:1994pre, Kidder:1995zr}.  Spin-precession leaves a direct
imprint on the GW signal and can be used to distinguish between
possible binary formation mechanisms. For example, while isolated
binaries formed in galactic fields are expected to have aligned
spins~\cite{Mapelli:2021for}, binaries formed via random encounters in
dense stellar clusters can have randomly oriented
spins~\cite{Mapelli:2021for}. To reliably extract this
astrophysical information from GW signals, accurate waveform
models~\cite{Varma:2019csw, Pratten:2020ceb, Ossokine:2020kjp,
Gamba:2021ydi,Estelles:2021gvs,Hamilton:2021pkf} and GW data analysis
methods~\cite{Farr:2014qka, Romero-Shaw:2020owr, Veitch:2014wba} that
capture the effects of spin-precession have been developed.

By contrast, orbital eccentricity leads to bursts of GW radiation at
every pericenter (point of closest approach)
passage~\cite{Peters:1963ux, Peters:1964zz}, which appear as orbital
timescale modulations of the GW amplitude and
frequency~\cite{Blanchet:2013haa}. The eccentricity of GW signals
carries information about the binary formation mechanism that is
complimentary to what can be learned from spin-precession alone. For
example, isolated galactic-field binaries are expected to become
circularized via GW emission~\cite{Peters:1963ux, Peters:1964zz}
before they enter the LIGO-Virgo frequency
band~\cite{Mapelli:2021for}. Because eccentric signals are considered
less likely for LIGO-Virgo, most analyses to date (e.g.\
Ref.~\cite{LIGOScientific:2021djp}) ignore eccentricity. However,
binaries formed via random encounters in dense clusters can merge
before they can circularize, thereby entering the LIGO-Virgo band with
a finite eccentricity~\cite{Mapelli:2021for}.  Similarly, in
hierarchical triple systems, the tidal effect of the tertiary can
excite periodic eccentricity oscillations of the inner
binary~\cite{Naoz:2016tri}, resulting in high-eccentricity mergers in
the LIGO-Virgo band~\cite{Martinez:2020tri}.

LIGO-Virgo observations can be used to ascertain whether the
assumptions of small eccentricity are valid, and to measure any
nonzero eccentricity that may be present. Therefore, eccentricity
measurements and/or upper limits from GW signals are highly sought
after, and several groups have already analysed the observed
signals to obtain information on
eccentricity~\cite{Romero-Shaw:2019itr, Romero-Shaw:2020thy,
Gayathri:2020coq, CalderonBustillo:2020fyi, CalderonBustillo:2020xms,
OShea:2021ugg, Gamba:2021gap,Romero-Shaw:2022xko}. As LIGO-Virgo, now
joined by KAGRA~\cite{KAGRA:2020tym}, continue to
improve~\cite{Aasi:2013wya}, and with next-generation ground-based
detectors expected in the
2030s~\cite{Punturo:2010zz,Hild:2010id,Evans:2016mbw,Reitze:2019iox},
future observations will enable stronger constraints on eccentricity.

The case for eccentric signals is stronger for the future space-based GW
observatory LISA, which will see the earlier inspiral phase of some of the BH
mergers observed by LIGO-Virgo~\cite{Sesana:2016ljz, Amaro-Seoane:2022rxf, Klein:2022rbf}, at
which point they may still have larger eccentricity. Furthermore, mergers of
supermassive black hole binaries observed by LISA may have significant
eccentricity if triple dynamics played a role in overcoming the final parsec
problem~\cite{Bonetti:2017dan}.  Finally, LISA will observe the mergers of
stellar mass compact objects with supermassive black holes, the so-called
extreme mass ratio inspirals (EMRIs). EMRIs are expected to primarily be formed
through dynamical capture leading to high eccentricities when entering the LISA
band~\cite{Amaro-Seoane:2022rxf}.

Driven by these observational prospects, there has been an increasing
effort to develop waveform models~\cite{Warburton:2011fk,
Osburn:2015duj, Cao:2017ndf, Liu:2019jpg, Ramos-Buades:2021adz,
Nagar:2021gss, Islam:2021mha, Liu:2021pkr, Memmesheimer:2004cv, Huerta:2014eca,
Tanay:2016zog, Cho:2021oai,Moore:2018kvz, Moore:2019xkm,
VanDeMeent:2018cgn, Chua:2020stf, Hughes:2021exa, Katz:2021yft,
Lynch:2021ogr,Klein:2021jtd}, gravitational self-force
calculations~\cite{Barack:2010tm, Akcay:2013wfa, Hopper:2012ty,
Osburn:2014hoa, vandeMeent:2015lxa, Hopper:2015icj, Forseth:2015oua,
vandeMeent:2016pee, vandeMeent:2017bcc, Munna:2020juq, Munna:2020som,
Munna:2022gio, Munna:2022xts}, numerical relativity (NR)
simulations~\cite{Hinder:2017sxy, Boyle:2019kee, Ramos-Buades:2022lgf,
Healy:2022wdn, Habib:2019cui, Huerta:2019oxn,Joshi:2022ocr}, and source parameter
estimation methods~\cite{LIGOScientific:2016ebw,Lower:2018seu, Ramos-Buades:2019uvh,
Romero-Shaw:2019itr, Romero-Shaw:2020thy, Gayathri:2020coq,
Ramos-Buades:2020eju, CalderonBustillo:2020fyi,
CalderonBustillo:2020xms, OShea:2021ugg, Gamba:2021gap,
Romero-Shaw:2022fbf, Romero-Shaw:2022xko, Knee:2022hth,
Bonino:2022hkj,Klein:2022rbf,Yang:2022tig} that include the effects of eccentricity. In addition
to these efforts, one important obstacle needs to be overcome in order
to reliably extract eccentricity from GW signals: Eccentricity is not
uniquely defined in general relativity~\cite{Blanchet:2013haa}, and
therefore most waveform models and simulations use custom internal
definitions that rely on gauge-dependent quantities like binary
orbital parameters or compact object trajectories. As a result, the
eccentricity inferred from GW signals can be riddled with ambiguity
and can even be incompatible between different
models~\cite{Knee:2022hth}. Such ambiguities propagate into any
astrophysical applications, including using eccentricity to identify
the binary formation mechanism. To resolve this problem, there is a
need for a standardized definition of eccentricity for GW
applications.

In addition to eccentricity, one needs two more parameters to fully
describe an eccentric orbit -- one describing the current position of the
bodies on the orbit relative to the previous pericenter passage and the other
describing the size of the orbit. Mean anomaly~\cite{Blanchet:2013haa,
Schmidt:2017btt, Islam:2021mha}, which is the fraction of the orbital period
(expressed as an angle) that has elapsed since the last pericenter passage, can
be used as the first parameter.~\footnote{While mean anomaly is the most
convenient choice in our experience, other choices for the second
parameter~\cite{Clarke:2022fma} like the ``true anomaly'' are also possible.}
The size of the eccentric orbit can be described, for example, by the
semi-major axis $a$ which is related to the orbital period $P$ as $a^3\propto
P^2$ for a Keplerian orbit. For general relativistic orbits, the orbital period
decreases as the binary inspirals, while the frequency increases. While the GW
frequency itself can be nonmonotonic for eccentric binaries, as we will discuss
in Sec.~\ref{sec:generalizing_fref}, one can construct an orbit-averaged
frequency that is monotonically increasing. Using such an orbit-averaged
frequency one can construct a one-to-one map between the orbit-averaged
frequency and the orbital period (and therefore the semi-major axis). Thus a
reference frequency like the orbital-averaged frequency can be used to describe
the size of the orbit.

A good definition of eccentricity should have the following features:
\begin{enumerate}[label=(\Alph*)]
\item To fully describe an eccentric orbit at a given reference
    frequency, two parameters are required: eccentricity and mean anomaly.
    Therefore, the definition should include both eccentricity and mean
    anomaly.
\item\label{item:gaugeinvariance} To avoid gauge ambiguities, eccentricity and mean anomaly should
  be defined using only observables at future null-infinity, like the
  gravitational waveform.
\item In the limit of large binary separation, the eccentricity
  should approach the Newtonian value, which is uniquely defined.
\item\label{item:ecc_range} The standardized definition should be applicable
  over the full range of allowed eccentricities for bound orbits ($0 - 1$). It
  should return zero for quasicircular inspirals and limit to one for marginally
  bound ``parabolic'' trajectories.
\item Because the eccentricity and mean anomaly vary during a binary's
  evolution, one must pick a point in the evolution at which to measure them.
  This is generally taken to be the point where the GW frequency reaches a
  certain reference value $\fref$ (typically
  20Hz~\cite{LIGOScientific:2021djp}). However, because eccentricity causes
  modulations in the GW frequency, the same $\fref$ can occur at multiple
  points. Therefore, the standardized definition should also prescribe how
  to select an unambiguous reference point for eccentric binaries.
\item As current GW detectors are only sensitive to frequencies above a certain
  $\fLow$ (typically 20Hz~\cite{LIGOScientific:2021djp}), when using
  time-domain waveforms, one typically discards all times below $\tLow$,
  chosen so that the GW frequency crosses 20Hz at $\tLow$. Once again,
  because the GW frequency is nonmonotonic, the standardized definition
  should prescribe how to select $\tLow$ for eccentric binaries.
\end{enumerate}

Additionally, the following features,
while not strictly required, can be important for practical applications:
\begin{enumerate}[label=(\alph*)]
\item\label{item:egeo_vs_egw} In the limit of large mass ratio, the
  eccentricity should approach the test particle eccentricity on a Kerr
  geodesic. Since the geodesic eccentricity is not uniquely defined, it is not
  strictly required that the standard definition of eccentricity matches the
  geodesic eccentricity defined in any particular coordinates. As described in
  Sec.~\ref{sec:egw_emri_limit}, the definition adopted in this work only
  approximately matches the geodesic eccentricity defined in the Boyer–Lindquist
  coordinates.
\item The eccentricity and mean anomaly computation should be computationally
  inexpensive and robust across binary parameter space and be applicable to a
  broad range of waveform models and NR simulations. Thus, most
  models/simulations can continue to rely on their internal eccentricity
  definitions as it is most convenient to conduct source parameter estimation
  using the internal definitions.  However, if the computation is cheap and
  robust, one can convert posterior samples from the internal definition to the
  standardized one as a postprocessing step, thus putting all models and
  simulations on the same footing.
\end{enumerate}

In this paper, we adopt a standardized eccentricity and mean anomaly definition
that meets all of the criteria in the first list of (required) features and
also satisfies the criteria in the second list of (desired but not strictly
required) features to a great extent.  Over the last few years, there have been
several attempts to standardize the definition of
eccentricity~\cite{Ramos-Buades:2019uvh, Islam:2021mha, Ramos-Buades:2019uvh,
Bonino:2022hkj}, or map between different definitions~\cite{Knee:2022hth}, but
these approaches either ignore mean anomaly, or do not have the correct limits
at large separation or large mass ratio~\cite{Ramos-Buades:2022lgf}. More
recently, Ref.~\cite{Ramos-Buades:2022lgf} introduced a new definition, that
has the correct limits, which we adopt in this work. We rigorously test and
demonstrate the robustness of our implementation on eccentric waveforms
spanning the full range of eccentricities and different origins: post-Newtonian
(PN) theory, NR, effective one body (EOB), and EMRIs.

While we focus on eccentric binaries without spin-precession for simplicity, we
include a discussion of how our methods can be extended to
spin-precessing eccentric systems.  In addition, we describe how $\fref$
and $\tLow$ should be generalized for eccentric binaries, along with a
discussion on the benefit of using dimensionless reference
points~\cite{Varma:2021csh}. Our computation is very cheap, and our
implementation can be used directly during source parameter estimation
or as a postprocessing step. We make our implementation publicly
available through an easy-to-use \python{} package \package{}.

This paper is organized as follows. In
Sec.~\ref{sec:defining_eccentricity}, we describe the standardized
eccentricity and mean anomaly definitions, along with a discussion of
how to generalize $\fref$ and $\fLow$. In
Sec.~\ref{sec:methods_to_locate_extrema}, we provide implementation
details, along with different choices for capturing the eccentricity
modulations in waveforms. In Sec.~\ref{sec:tests}, we demonstrate the
robustness of our implementation on waveforms of different origins and
over the full range of eccentricities. We finish with some concluding
remarks in Sec.~\ref{sec:conclusion}.

%==========================================================================
\section{Defining eccentricity}
\label{sec:defining_eccentricity}

\subsection{Notation and conventions}
The component masses of a binary are denoted as $m_1$ and $m_2$, with $m_1 \geq
m_2$, total mass $M=m_1+m_2$, and mass ratio $q=m_1/m_2\geq1$. The
dimensionless spin vectors of the component objects are denoted as
$\bm{\chi}_1$ and $\bm{\chi}_2$, and have a maximum magnitude of 1. For
binaries without spin-precession, the direction of the orbital angular momentum $\bm{L}$
is fixed, and is aligned to the $z$-axis by convention. For these binaries, the
spins are constant and are aligned or anti-aligned to $\bm{L}$, meaning that
the only nonzero spin components are $\chi_{1z}$ and $\chi_{2z}$.

The plus ($h_+$) and cross ($h_{\into}$) polarizations of GWs can be
conveniently represented by a single complex time series $\h = h_{+} -i \,
h_{\into}$. The complex waveform on a sphere can be decomposed into a sum of
spin-weighted spherical harmonic modes $\hlm$, so that the waveform along any
direction $(\iota, \varphi_0)$ in the binary's source frame is given by
\begin{equation}
\label{eq:h}
\h(t, \iota, \varphi_0) =
\sum_{\ell=2}^{\ell=\infty}\sum_{m=-\ell}^{m=\ell} \hlm(t) ~
{}_{-2}Y_{\ell m}(\iota,\, \varphi_0),
\end{equation}
where $\iota$ and $\varphi_0$ are the polar and azimuthal angles on the sky in
the source frame, and ${}_{-2}Y_{\ell m}$ are the spin$=-2$ weighted spherical
harmonics. Unless the total mass and/or distance are explicitly specified, we
work with the waveform at future null-infinity scaled to unit total mass and
distance for simplicity. We also shift the time array of the waveform such that
$t=0$ occurs at the peak of the amplitude of the dominant $\dm$
mode.~\footnote{When generalizing to spin-precessing binaries, this should be
replaced by the total waveform amplitude, defined in Eq.~(5) of
Ref.~\cite{Varma:2019csw}.} We note, however, that the implementation in
\package{} handles waveforms in arbitrary units and time conventions.

\subsection{Eccentricity definitions used in PN, EOB, self-force and NR}
Because eccentricity is not uniquely defined in general relativity, a wide
variety of definitions of eccentricity exists. At Newtonian order, eccentricity
can be uniquely defined as~\cite{goldstein2002classical}.
\begin{equation}
\label{eq:e_Newtonian}
  \eNewtonian = \frac{r^{\text{a}}-r^{\text{p}}}{r^{\text{a}}+r^{\text{p}}},
\end{equation}
where $r^{\text{a}}$ and $r^{\text{p}}$ are the separations at apocenter (point
of furthest approach) and pericenter (point of closest approach), respectively.
Starting at 1PN order, the Keplerian parametrization can be extended to the
so-called \emph{quasi-Keplerian} parametrization where three different
eccentricity parameters are defined, the radial $e_r$, temporal $e_t$ and
angular $e_{\phi}$ eccentricities, each of which has the same Newtonian
limit~\cite{Blanchet:2013haa}. These quantities can be defined in terms of the
conserved energy and angular momentum, but depend on the gauge
used~\cite{Memmesheimer:2004cv}.

The Bondi energy and angular momentum of a
binary can be accessed from the metric at future null-infinity and are (nearly)
gauge invariant. One might therefore hope to formulate a definition of
eccentricity based purely on these two quantities that satisfies all of our
requirements. Unfortunately, this will be
challenging for the following reasons: Suppose we define eccentricity as some
function $e(E, J)$ of the Bondi energy $E$ and angular momentum $J$. The
equation $e=0$ will generically define a 1-dimensional subset of the
$(E, J)$-plane, which would be shared by all quasicircular inspirals. However, the
track followed by a quasicircular binary through the $(E, J)$-plane depends on
the mass ratio and the spins (e.g. see
Ref.~\cite{Ossokine:2017dge}). Consequently, a definition of eccentricity based
purely on $E$ and $J$ cannot assign zero eccentricity to all quasicircular
inspirals, i.e. it cannot satisfy requirement \ref{item:ecc_range} above.  

One might further hope to overcome this by adding an explicit dependence on the
mass ratio and spins to the definition. But this cannot account for the fact that different inspiral models
will, in general, still not agree on the location of the $e=0$ locus.
Consequently, whatever reference model is chosen as the basis for the
definition, it will be unable to assign zero eccentricity to quasicircular
inspirals produced by all models.  This is made worse by the fact that the
$e=0$ locus represents the edge of the allowable range of $E$ and $J$; if the
values of $E$ and $J$ for some model lie outside the range of the reference
model (e.g. see ef.~\cite{Ramos-Buades:2022lgf}), analytically inverting the
relationship with $e$ would assign a complex value to $e$.
	
Finally, one might hope to cure this behaviour by basing the definition
of $e$ on $E-E_{\text{qc}}$, where $E_{\text{qc}}$ is the energy of the quasicircular
counterpart with same angular momentum (and mass-ratio and spins) as the model
being measured.  However, $E_{\text{qc}}$ is not something that can be inferred
from observables at null-infinity (violating
requirement~\ref{item:gaugeinvariance}).  Furthermore, $E_{\text{qc}}$ may not be
straightforward to obtain in some models (e.g. numerical relativity), unless
one relies on a reference model, which comes with the problems noted above.
%Moreover, because relations between
%$E$, $J$, binary parameters and eccentricity are only known approximately (e.g. to a
%certain PN order), $e=0$ would only approximately correspond to a
%quasicircular inspiral. 
We, therefore, do not take this approach. Similar
objections arise with gauge invariant definitions of eccentricity based on the
radial and azimuthal periods, as are commonly used to facilitate gauge
invariant comparisons between self-force and PN results for eccentric
orbits~\cite{Barack:2011ed, Akcay:2015pza, Akcay:2016dku}.

In the EOB formalism, initial conditions for the dynamics are prescribed in
terms of an eccentricity parameter defined within the quasi-Keplerian
parameterization~\cite{Hinderer:2017jcs, Chiaramello:2020ehz, Khalil:2021txt,
Nagar:2021gss, Ramos-Buades:2021adz}.  Thus, the gauge dependency of the
eccentricity parameter also extends to the EOB
waveforms~\cite{Ramos-Buades:2021adz, Nagar:2021gss}. In self-force
calculations for EMRIs, one typically uses an eccentricity definition based on
the turning points of the underlying geodesics~\cite{Warburton:2011fk,
    Osburn:2015duj, VanDeMeent:2018cgn, Chua:2020stf, Hughes:2021exa,
Katz:2021yft, Lynch:2021ogr}.  This is inherently dependent on the coordinates
used for the background spacetime, and picks-up further gauge ambiguities at
higher orders in the mass ratio. For NR waveforms, the compact object
trajectories are used to define eccentricity, typically by fitting to
analytical PN (or Newtonian) expressions~\cite{Buonanno:2010yk, Mroue:2012kv,
Ramos-Buades:2018azo, Ciarfella:2022hfy}. This also inherently depends on the
gauge employed in the simulations.

\subsection{Defining eccentricity using the waveform}
\label{sec:defining_eccentricity_using_waveform}

A more convenient definition of eccentricity that can be straightforwardly
applied to waveforms of all origins was proposed in Ref.~\cite{Mora:2002gf}:
\begin{equation}
  \label{eq:mora_will}
  \eOrb(t) = \frac{\sqrt{\omegaOrbP(t)} -
    \sqrt{\omegaOrbA(t)}}{\sqrt{\omegaOrbP(t)} + \sqrt{\omegaOrbA(t)}},
\end{equation}
where $\omegaOrbP(t)$ is an interpolant through the orbital frequency
$\omegaOrb(t)$ evaluated at pericenter passages, and likewise for
$\omegaOrbA(t)$ at apocenter passages. Because eccentricity causes a burst of
radiation at pericenters, the times corresponding to pericenters are identified
as local maxima in $\omegaOrb(t)$, while apocenters are identified as local
minima. Eq.~(\ref{eq:mora_will}) was used, for example, in
Ref.~\cite{Lewis:2016lgx} to analyze generic spin-precessing and eccentric binary BH
waveforms. Unfortunately, because $\omegaOrb$ is computed using the compact
object trajectories, Eq.~\eqref{eq:mora_will} is also susceptible to gauge
choices, especially for NR simulations.

Nevertheless, Eq.~\eqref{eq:mora_will} has the important quality that it can be
applied to waveforms of all origins. Furthermore, Eq.~\eqref{eq:mora_will} has
the correct Newtonian limit. This is easily seen using Kepler's second law
$\omegaOrb \propto 1/r^2$, where $r$ is the binary
separation~\cite{goldstein2002classical, KeplersLaw}. Using this relation in
Eq.~(\ref{eq:mora_will}), one finds that $\eOrb$ matches $\eNewtonian$ from
Eq.~\eqref{eq:e_Newtonian}.

The main limitation of Eq.~\eqref{eq:mora_will} is that $\omegaOrb$ is
gauge-dependent. To remove such dependence, one must turn to the waveform at
future null-infinity, which is where our detectors are approximated to
be with respect to the source. The emitted GWs can be obtained at future
null-infinity, for example, by evolving Einstein's equations along null
slices~\cite{Winicour:2008vpn, Moxon:2020gha, Reisswig:2006nt, Reisswig:2009rx, Reisswig:2012ka, Taylor:2013zia, Barkett:2019uae}. While the waveform at future
null-infinity is unique up to Bondi-Metzner-Sachs (BMS) transformations,
this freedom can be fixed using BMS charges~\cite{Mitman:2022kwt}. In the rest of
this paper, we assume this freedom has been fixed, but our method can also be
applied to waveforms specified in any given frame.

\begin{figure*} \centering
\includegraphics[width=0.95\textwidth,trim=0 5 0 6]{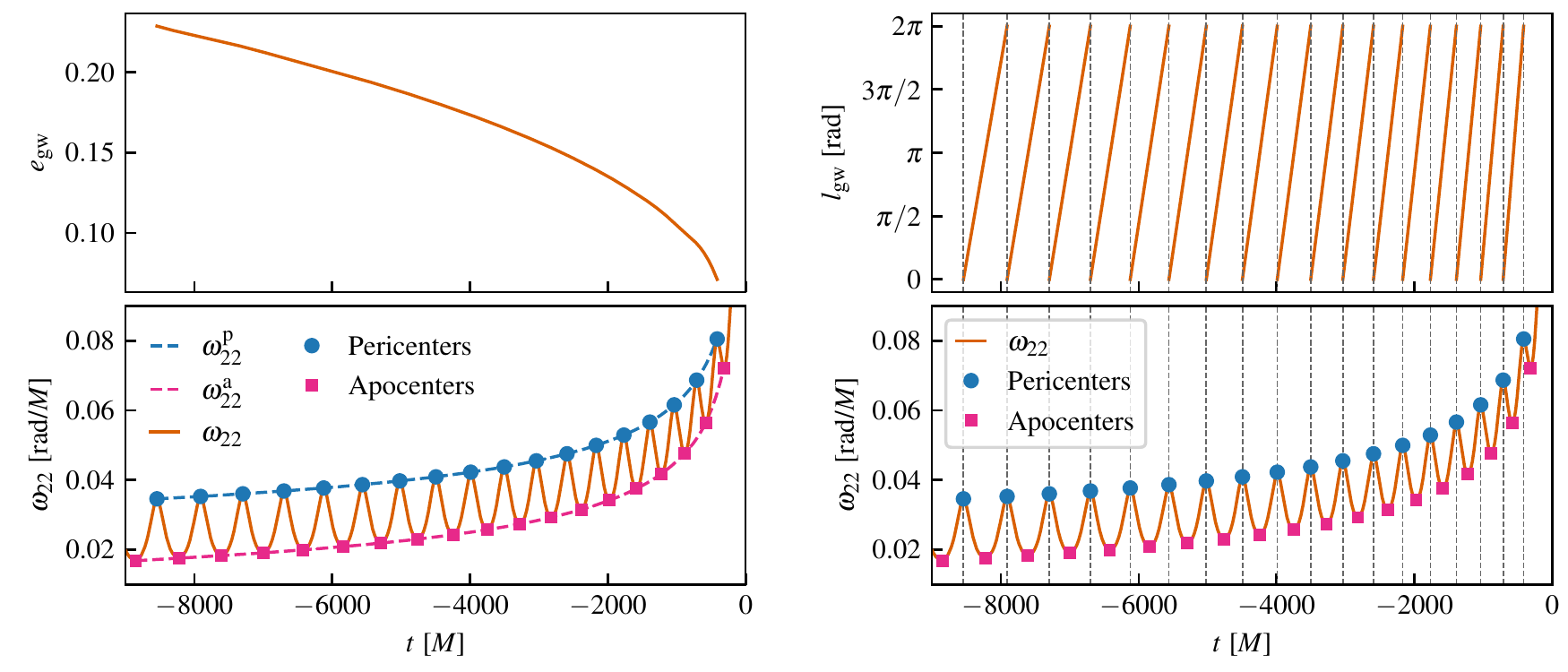}
\caption{
Eccentricity and mean anomaly measured using the waveform from an equal-mass
nonspinning eccentric NR simulation
(\texttt{SXS:BBH:2312}~\cite{SXSCatalog,Islam:2021mha}). {\itshape Left:} Time
evolution of the eccentricity $\egw$ (upper panel) and frequency of the $\dm$
waveform mode $\omegatwotwo$ (lower panel). $\omegaP(t)$ and $\omegaA(t)$ are
interpolants through $\omegatwotwo(t)$ evaluated at the pericenters (\cPeri{}
\markPeri{}s) and apocenters (\cApo{} \markApo{}s), respectively.  Eq.~\eqref{eq:eGW} is used to
compute $\egw(t)$ given $\omegaP(t)$ and $\omegaA(t)$.  {\itshape Right:} Time
evolution of the mean anomaly $\lgw$ (upper panel) and $\omegatwotwo$ (lower
panel). The vertical dashed \cVline{} lines denote the pericenter times.
$\lgw(t)$ grows linearly in time from $0$ to $2\pi$ between successive
pericenters (Eq.~\eqref{eq:mean_anomaly_definition}).}
\label{fig:ecc_definition}
\end{figure*}

For a gauge-independent definition of eccentricity, we seek an analogue of
Eq.~\eqref{eq:mora_will} that only depends on the waveform $\hlm$. The simplest
possible generalization~\cite{Ramos-Buades:2019uvh, Islam:2021mha,
Ramos-Buades:2021adz, Bonino:2022hkj} is to replace the
trajectory-dependent orbital frequency $\omegaOrb(t)$ in
Eq.~\eqref{eq:mora_will} with the frequency of the dominant $\dm$ mode
$\omegatwotwo(t)$:
\begin{equation}
  \label{eq:e_omegatwotwo}
  \etwotwo(t) = \frac{\sqrt{\omegaP(t)} -
    \sqrt{\omegaA(t)}}{\sqrt{\omegaP(t)} + \sqrt{\omegaA(t)}},
\end{equation}
where $\omegaP(t)$ and $\omegaA(t)$ are interpolants through $\omegatwotwo(t)$
evaluated at pericenters and apocenters, respectively. $\omegatwotwo$ is
obtained from $\htwotwo$ as follows:
\begin{align}
\htwotwo(t) = \Atwotwo(t)\,e^{- i \phitwotwo(t)}, \\
\omegatwotwo(t) = \frac{\dd \phitwotwo(t)}{\dd t},
\label{eq:omega_22}
\end{align}
where $\Atwotwo$ is the amplitude and $\phitwotwo$ the phase of $\htwotwo$.

In Eq.~\eqref{eq:e_omegatwotwo}, the pericenter and apocenter times can
be chosen to correspond to local maxima and minima, respectively, in
$\omegatwotwo(t)$. This procedure is illustrated in the bottom-left panel of
Fig.~\ref{fig:ecc_definition}. It is not guaranteed that the local extrema of
$\omegatwotwo$ coincide with the local extrema of $\omegaOrb$. Instead, we can
\emph{define} the local extrema of $\omegatwotwo$ to correspond to pericenters
and apocenters. Other choices for assigning pericenter/apocenter times and
their impact on the eccentricity will be discussed in
Sec.~\ref{sec:methods_to_locate_extrema}.

Because of its simplicity and gauge-independent nature,
Eq.~\eqref{eq:e_omegatwotwo} has been applied to parameterize eccentric
waveforms as well as GW data analysis~\cite{Ramos-Buades:2019uvh,
Islam:2021mha, Ramos-Buades:2021adz, Bonino:2022hkj}. However, as shown in
Ref.~\cite{Ramos-Buades:2022lgf}, this definition of eccentricity does not have
the correct Newtonian limit at large separations. In particular, in the small
eccentricity limit at Newtonian order, one obtains~\cite{Ramos-Buades:2022lgf}:
\begin{equation}
  \lim_{e_t \rightarrow 0 }e^{\text{0PN}}_{\omegatwotwo} = \frac{3}{4} e_t + \mathcal{O}(e_t^3),
  \label{eq:eqE3}
\end{equation}
where $e_t$ is the temporal eccentricity used in PN theory, which matches the
Newtonian eccentricity at Newtonian order~\cite{Blanchet:2013haa}.

This discrepancy can be resolved by using the following
transformation~\cite{Ramos-Buades:2022lgf}
\begin{equation}
\label{eq:eGW}
  \egw = \cos(\Psi/3) - \sqrt{3} \, \sin(\Psi/3),
\end{equation}
where
\begin{equation}
  \label{eq:psi}
  \Psi = \arctan\left(\frac{1 - \etwotwo^2}{2\,\etwotwo}\right).
\end{equation}
Eq.~\eqref{eq:eGW} has the correct Newtonian limit over the full range of
eccentricities~\cite{Ramos-Buades:2022lgf}, and we adopt this definition in
this work. As we will show in Sec.~\ref{sec:egw_emri_limit}, $\egw$ also
approximately matches the geodesic eccentricity in the extreme mass ratio
limit, while $\etwotwo$ does not.

The top-left panel of Fig.~\ref{fig:ecc_definition} shows an example
evaluation of $\egw(t)$ for an NR simulation produced using the
Spectral Einstein Code~\cite{Boyle:2011gg, SXSCatalog} (\SpEC{}),
developed by the Simulating eXtreme Spacetimes (SXS)
collaboration~\cite{SXSWebsite}. As expected, $\egw$ monotonically
decreases as the binary approaches the merger ($t=0$). However, while
the waveform itself covers the full range of times shown, $\egw(t)$
does not. This is because $\egw(t)$ depends on the $\omegaP(t)$ and
$\omegaA(t)$ interpolants in Eq.~\eqref{eq:e_omegatwotwo}, which do
not span the full time range, as shown in the bottom-left panel of
Fig.~\ref{fig:ecc_definition}. $\omegaP(t)$ is only defined between
the first and last available pericenters, and $\omegaA(t)$ is only defined
between the first and last available apocenters. Therefore, the first available
time for $\egw(t)$ is the maximum of the times of the first pericenter
and first apocenter. Similarly, the last available time for $\egw(t)$
is the minimum of the times of the last pericenter and last apocenter.

Furthermore, we find that $\egw(t)$ near the merger can become nonmonotonic,
which is not surprising as it becomes hard to define an orbit in this
regime. To avoid this nonmonotonic behavior, we discard the last two orbits of
the waveform before computing $\egw$. As a result, the last available time for
$\egw$ is the minimum of the times of the last pericenter and last apocenter in
the remaining waveform, which falls at about two orbits before the peak
amplitude. In addition, to successfully build the $\omegaP(t)$ and $\omegaA(t)$
interpolants in Eq.~\eqref{eq:e_omegatwotwo}, we require at least two orbits in
the remaining waveform. Therefore, the full waveform should include at least
$\roughly 4-5$ orbits to reliably compute $\egw$.

\subsubsection{Extending to spin-precessing and frequency-domain waveforms}
\label{sec:extending_to_precessing_and_FD}
Eqs.~\eqref{eq:e_omegatwotwo} and \eqref{eq:eGW} use only the $\dm$ mode as it
is the dominant mode of radiation~\cite{Varma:2016dnf, Varma:2014jxa,
Capano:2013raa}, at least for binaries without spin-precession in which the direction of
the orbital angular momentum is fixed (taken to be along $\hat{z}$ by
convention). On the other hand, for spin-precessing binaries, the orbital angular
momentum direction varies, and the power of the $\dm$ mode leaks into the other
$\ell=2$ modes, meaning that there need not be a single dominant mode of
radiation. For this reason, we restrict ourselves to binaries without spin-precession in
this work. We expect that our method can be generalized to spin-precessing binaries
by using $\htwotwo$ in the coprecessing frame~\cite{Boyle:2011gg,
Schmidt:2010it, OShaughnessy:2011pmr}, which is a non-inertial frame that
tracks the binary's spin-precession so that $\hat{z}$ is always along the
instantaneous orbital angular momentum.  Alternatively, one could replace
$\omegatwotwo$ in Eq.~\eqref{eq:e_omegatwotwo} with a frame-independent angular
velocity~\cite{Boyle:2013nka} that incorporates information from all available
waveform modes.

We also restrict ourselves to time-domain waveforms in this work. One main
difficulty for frequency-domain waveforms~\cite{Moore:2018kvz,Moore:2019xkm}
is the identification of the frequencies at which pericenters and apocenters
occur. This is complicated by the fact that even for the $\dm$ mode,
eccentricity excites higher harmonics that make it difficult to identify local
extrema in the frequency domain (see e.g.\ Fig.~3 of Ref.~\cite{Moore:2018kvz}).
Alternatively, one could simply apply an inverse Fourier transform to first
convert the frequency-domain waveform to time-domain, although this can be
computationally expensive for long signals.

\subsection{Defining mean anomaly using the waveform}
\label{sec:defining_meanano_using_waveform}

To fully describe an eccentric orbit at a given reference frequency, two
parameters are required: eccentricity and mean anomaly~\cite{Blanchet:2013haa,
Schmidt:2017btt, Islam:2021mha}, which is the fraction of the orbital period
(expressed as an angle) that has elapsed since the last pericenter
passage. Similar to $\egw$, we seek a definition of mean anomaly that depends
only on the waveform at future null-infinity. This can be achieved by
generalizing the Newtonian definition of mean anomaly to~\cite{Schmidt:2017btt,
Ramos-Buades:2019uvh, Ramos-Buades:2022lgf, Islam:2021mha}
\begin{equation}
\label{eq:mean_anomaly_definition}
  \lgw(t) = 2\pi \, \frac{t - \tP_{i}}{\tP_{i+1} - \tP_{i}},
\end{equation}
defined over the interval $\tP_{i} \leq t < \tP_{i+1}$ between any two
consecutive pericenter passages $\tP_i$ and $\tP_{i+1}$. $\lgw$ grows linearly
in time over the range $[0, 2\pi)$ between $t=\tP_i$ and $t=\tP_{i+1}$. In
Newtonian gravity, the period of the orbit $T = \tP_{i+1} - \tP_{i}$ remains
constant, while in general relativity, radiation reaction cause $T$ to decrease
over time, making $\lgw(t)$ a stepwise linear function whose slope increases as
as the binary approaches the merger. As the times corresponding to pericenter
passages are already determined when calculating $\egw$, computing $\lgw$ is
straightforward. This procedure is illustrated in the right panel of
Fig.~\ref{fig:ecc_definition}.

We stress that the mean anomaly cannot be absorbed into a time or phase
shift~\cite{Islam:2021mha}, and is instead an intrinsic property of the binary
like the component masses, spins and $\egw$. This can be seen from the
bottom-right panel of Fig.~\ref{fig:ecc_definition}, showing $\omegatwotwo(t)$.
Consider the first pericenter occurring at $t\simeq-8500M$, for which $\lgw=0$.
First, because $\omegatwotwo$ is insensitive to phase shifts, one cannot apply
a phase shift to change the mean anomaly at $t\simeq-8500M$ away from $\lgw=0$.
Similarly, one cannot apply a time shift so that the mean anomaly at
$t\simeq-8500M$ is changed, without simultaneously also changing the frequency
at that time (because the time shift also applies to $\omegatwotwo(t)$). In
other words, to change the mean anomaly at a fixed time before the merger,
one also needs to change the frequency at a fixed time before the merger, which
results in a different physical system. Ignoring mean anomaly in
waveform models and/or parameter estimation can result in systematic biases in
the recovered source parameters~\cite{Islam:2021mha, Clarke:2022fma,
Ramos-Buades:2023yhy}.

\subsection{Generalizing the reference frequency $\fref$}
\label{sec:generalizing_fref}

Binary parameters like the component spin directions, and orientation with respect
to the observer, as well as eccentricity and mean anomaly, can vary during a
binary evolution. Therefore, when measuring binary parameters from a GW signal,
one needs to specify at which point of the evolution the measurement should be
done. This is typically chosen to be the point at which the GW frequency
crosses a reference frequency $\fref$, with a typical choice of
$\fref=20$Hz~\cite{LIGOScientific:2021djp} as that is approximately where the
sensitivity band of current ground-based detectors begins.

\begin{figure}
\centering
\includegraphics[width=\columnwidth,trim=0 5 0 4]{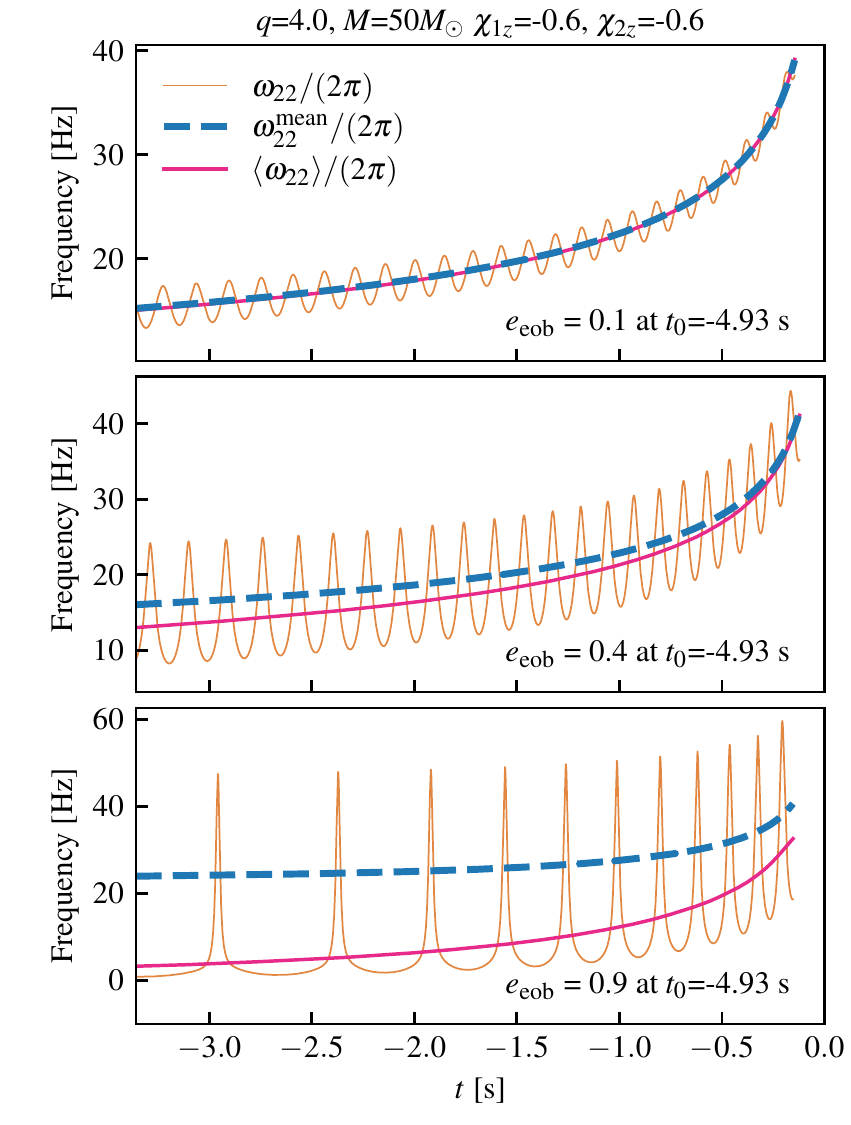}
\caption{
Different methods to construct a monotonically increasing frequency to replace
$\omegatwotwo(t)$, in order to set the reference frequency $\fref$ for
eccentric binaries. We consider two different approaches: (i) $\omegaMean(t)$,
the mean of $\omegaP(t)$ and $\omegaA(t)$, and (ii) $\avgOmega(t)$, an
interpolant through the orbit averaged $\omegatwotwo$
(Eq.~\eqref{eq:mean_motion}).
We show \SEOB{} waveforms with three different eccentricities; the binary
parameters are given in the figure text. While the two approaches agree for
small eccentricities, they deviate significantly at large eccentricities. We
adopt $\avgOmega(t)$ as it captures the correct frequency scale in
an orbit-averaged sense (Sec.~\ref{sec:generalizing_fref}).
}
\label{fig:omega22_average}
\end{figure}

For quasicircular binaries without spin-precession, the GW frequency increases
monotonically, and $\fref$ can be uniquely associated with a reference time
$\tref$. For spin-precessing, quasicircular binaries, while $\omegatwotwo$ in the
inertial frame can be nonmonotonic, one can use the frequency computed in the
coprecessing frame, which is always monotonically
increasing~\cite{Boyle:2011gg, Varma:2019csw}. Unfortunately, no such frame
exists for eccentric binaries, and $\omegatwotwo$ becomes nonmonotonic if
eccentricity is sufficiently high (see Fig.~\ref{fig:ecc_definition}).

Therefore, unique specification of a reference point via a frequency $\fref$
requires a generalization of $\omegatwotwo$ that is monotonically increasing,
and approaches $\omegatwotwo$ in the quasicircular limit.  In the following we
discuss two different ways to accomplish this and point out why the second is
superior.

\subsubsection{Mean of $\omegaP(t)$ and $\omegaA(t)$}

A simple method to compute a monotonically increasing frequency for eccentric
binaries is to take the mean of the interpolants through the frequencies at
pericenters ($\omegaP(t)$) and apocenters ($\omegaA(t)$), both of which are
monotonically increasing functions of time:
\begin{equation}
\label{eq:omegaAvg_peri_mean}
\omegaMean(t) = \frac{1}{2}\left[\omegaP(t) + \omegaA(t)\right],
\end{equation}
with the reference time defined as $\omegaMean(\tref) = 2\pi \fref$.

As $\omegaP(t)$ and $\omegaA(t)$ are already constructed when computing $\egw$,
there is no additional computational cost. Furthermore, as $\omegaP$ and
$\omegaA$ approach $\omegatwotwo$ in the quasicircular limit, so does
$\omegaMean$. This method was used to set the reference frequency in
Ref.~\cite{Bonino:2022hkj}. Figure~\ref{fig:omega22_average} shows examples of
$\omegaMean(t)$ for waveforms produced using the
\SEOB{}~\cite{Ramos-Buades:2021adz} eccentric EOB model, for three different
values of the model's internal eccentricity parameter $\eEOB$, defined at a
time $\tStart=-4.93$~s before the peak amplitude.

\subsubsection{Orbit averaged $\omegatwotwo$}

Alternatively, one can use the orbit average of $\omegatwotwo$ in fixing the
reference point. Between any two consecutive pericenters $\tP_i$ and
$\tP_{i+1}$ we define
\begin{align}
  \avgOmega_{i}^{\text{p}} & = \frac{1}{\tP_{i+1} -
                  \tP_{i}} \, \int_{\tP_{i}}^{\tP_{i+1}}\omegatwotwo(t)\,\dd t
                  \nonumber \\
\label{eq:mean_motion}
& = \frac{\phitwotwo(\tP_{i+1}) - \phitwotwo(\tP_{i})}{\tP_{i+1} - \tP_{i}},
\end{align}
and associate $\avgOmega_i^\text{p}$ with the midpoint between $\tP_{i}$
and $\tP_{i+1}$:
\begin{equation}
  \avgTP_{i} = \frac{1}{2}\left(\tP_i + \tP_{i+1}\right).
\end{equation}

Applying this procedure to all consecutive pairs of pericenter times, we obtain
the set $\big\{\,\left(\avgTP_i, \avgOmega_i^\text{p}\right)\,\big\}$.
Similarly, using all consecutive pairs of apocenter times $\tA_i$ and
$\tA_{i+1}$, we obtain the set
$\left\{\,\big(\avgTA_i,\avgOmega_i^{\text{a}}\big)\,\right\}$.  Taking the
union of these two datasets, we build a cubic spline interpolant in
time to obtain $\avgOmega(t)$.

The resulting orbit averaged frequency $\avgOmega(t)$ is also monotonically
increasing and reduces to $\omegatwotwo(t)$ in the quasicircular limit.  The
reference time associated with a reference frequency is now determined via
\begin{equation}\label{eq:mean_motion_fref} \avgOmega(\tref)=2\pi \fref.
\end{equation} This method was used in
Refs.~\cite{Ramos-Buades:2023yhy,Ramos-Buades:2022lgf}. Compared to
$\omegaMean(t)$, $\avgOmega(t)$ has the added costs of computing orbit averages
and constructing a new interpolant.  The orbit averages are very cheap to
compute as they can be written in terms of phase differences
(Eq.~\eqref{eq:mean_motion}). The cost of the interpolant scales with the
number of orbits but it is generally also cheap to construct.

Figure~\ref{fig:omega22_average} also shows $\avgOmega(t)$ for the same \SEOB{}
waveforms. While $\omegaMean(t)$ and $\avgOmega(t)$ agree at small
eccentricities, they deviate significantly at large eccentricities. Unlike
$\omegaMean(t)$, $\avgOmega(t)$ has the additional property, albeit only in an
orbit-averaged sense, that at the time $\tref$ where $\avgOmega(\tref) = 2 \pi
\fref$, one GW cycle occurs over a time scale of $1/\fref$. This also
explains why for the high eccentricity case in Fig.~\ref{fig:omega22_average}
(bottom panel), $\avgOmega$ follows the general trend of $\omegatwotwo$ more
closely than $\omegaMean$. For these reasons, we will adopt $\avgOmega$ and
Eq.~\eqref{eq:mean_motion_fref} in the rest of the paper.

\subsection{Selecting a good reference point}
\label{sec:selecting_reference_point}

Given a reference frequency $\fref$, Sec.~\ref{sec:generalizing_fref} describes
how that can be used to pick a reference time, $\tref$, in the binary's
evolution.  Another important choice is what frequency to use for $\fref$. Most
current analyses for ground-based detectors use $\fref=20$
Hz~\cite{LIGOScientific:2021djp}, but we argue that this may not be suitable
for eccentric binaries. Setting $\fref=20$ Hz means that the reference time is
chosen to be the point where the observed GW frequency (or its orbit average)
at the detector crosses 20 Hz. However, the observed GW signals are redshifted
because of cosmological expansion, and the observed GW frequency depends on the
distance between the source and detector. Two identical binaries placed at
different distances would therefore reach an observed frequency of 20 Hz at
different points in their evolution. Because the eccentricity varies during the
evolution, the measured eccentricities for these binaries will be different
when they reach $\fref=20$ Hz at the detector! This is particularly problematic
for applications like constraining the astrophysical distribution of
eccentricities of GW sources, as the same source can be mistaken to have two
different eccentricities.

All binary parameters that vary during a binary's evolution, like spin
directions, could be prone to this problem.  However, because spin tilts vary
over spin-precession time scales spanning many orbits, this has not been a
significant issue so far when constraining the astrophysical spin
distribution~\cite{LIGOScientific:2021psn}, with the exception of
Ref.~\cite{Varma:2021xbh} where this effect was found to be important when
modeling the full 6D spin distribution. Eccentricity, on the other hand, can
change rapidly on an orbital time scale, especially in the late stages near the
merger (see Fig.~\ref{fig:ecc_definition}).

One way to avoid this problem is to use the GW frequency defined in the source
frame instead of the detector frame. However, this requires assuming a
cosmological model to compute the redshift between the two frames.
This can be problematic for applications like independently extracting
cosmological parameters like the Hubble parameter from GW
signals~\cite{LIGOScientific:2021aug}. Alternatively, one can use a
dimensionless reference frequency $M \fref$ or time $\tref/M$ as proposed by
Ref.~\cite{Varma:2021csh}, where $M$ is the total mass in the detector frame.
Both of these choices have the benefit of not depending on the distance to the
source as the total mass measured in the detector frame is also redshifted and
exactly cancels out the redshift of $\fref$ and $\tref$.
Ref.~\cite{Varma:2021csh} proposed reference points of $\tref/M = -100$ (where
$t=0$ is at the peak of the GW amplitude) and $M\fref=6^{-3/2}$ (the
Schwarzschild inner-most-circular-orbit (ISCO) frequency), as these always
occur close to the merger for comparable mass binaries, and certain spin
parameters like the orbital-plane spin angles are best measured near the
merger. For measuring eccentricity, an earlier dimensionless time or frequency
may be more appropriate, as eccentricity can be radiated away before the binary
approaches merger.

A more straightforward approach could be to set the reference point at a fixed
number of orbits before a fixed dimensionless time ($\tref/M$) or dimensionless
orbit-averaged frequency ($M\avgOmega$). Here, we define one orbit as the
period between two pericenter passages, as measured from the waveform. As the
number of orbits defined with respect to a dimensionless time/frequency is also
unaffected by the redshift, this serves the same purpose as a dimensionless
time/frequency. The number of orbits also scales more naturally to EMRI systems,
while dimensionless time/frequency may not. A similar approach was recently
adopted by Ref.~\cite{Romero-Shaw:2022fbf}.

Another advantage of using a fixed number of orbits before a dimensionless
time/frequency is that by using pericenters to define the number of orbits, we
can always measure eccentricity at a fixed mean anomaly of $\lgw=0$. This can
make it simpler to report posteriors for eccentric GW signals by reducing the
dimensionality by one. Similarly, this can make it easier to connect GW
observations to astrophysical predictions for GW populations, as the
predictions would just need to be made at a single mean anomaly value. However,
we stress that mean anomaly would still need to be included as a parameter in
waveform models and parameter estimation, and it is only when computing the
eccentricity from the waveform predictions in postprocessing that this
simplification occurs.

To summarize, while the most appropriate choice will need to be determined by
analyzing eccentric GW signals in a manner similar to
Ref.~\cite{Varma:2021csh}, we propose that the reference point be chosen to be
a fixed number of orbits (e.g.\ 10) before a fixed dimensionless time
(e.g.\ $\tref/M=-100$) or a fixed dimensionless orbit-averaged frequency
(e.g.\ $M\avgOmega=2\pi \, 6^{-3/2}$, the Schwarzschild ISCO frequency). While
not all GW signals will enter the detector frequency band with $\sim 10$ orbits
to go before the merger, this can be achieved by always generating GW templates
with at least 10 orbits when analyzing the GW signals. One important question
that remains is whether using a reference point that falls outside the detector
band leads to systematic biases or complications during parameter estimation.
We expect that as long as the number of orbits by which the reference point
falls outside the band is small, such effects should be small, but we leave
this investigation to future work.

\subsection{Truncating eccentric time domain waveforms}
\label{sec:generalizing_fLow}

GW detectors are most sensitive over certain frequency bands ($\sim 20$ Hz to
$\sim 10^3$ Hz for LIGO-Virgo), and waveform predictions need to include all physical GW frequencies present in this region. For frequency
domain waveform models this is achieved by evaluating the model starting at
initial frequency $\fLow=20$ Hz. On the other hand, time-domain waveform models
need to be evaluated starting at an initial time $\tLow$, chosen so that the GW
signal at earlier times does not contain any frequencies above $\fLow$. In
other words, the part of the time domain waveform that is not included
($t<\tLow$) does not contribute to the GW signal in the detector frequency
band.

\begin{figure}
\centering
\includegraphics[width=\columnwidth]{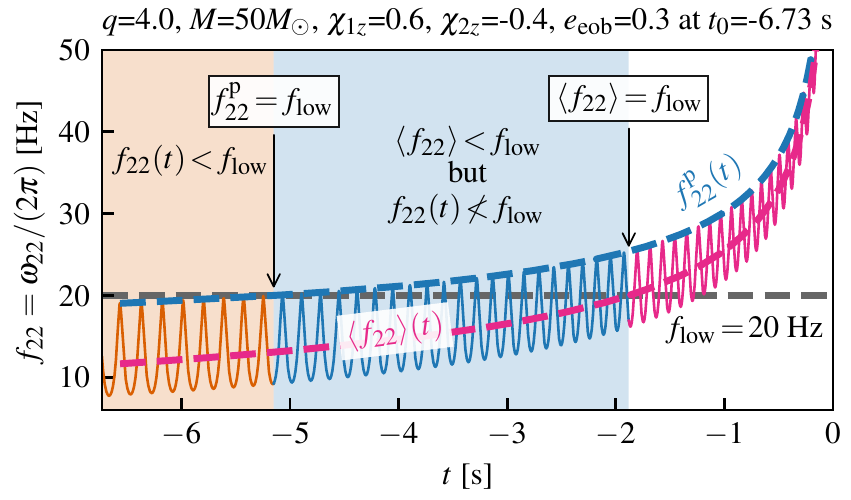}
\caption{ How to truncate time domain eccentric waveforms while
retaining all frequencies above $\fLow=20$ Hz. The orange, blue and
pink curves show different sections of $\omegatwotwo(t)$ for an
eccentric \SEOB{} waveform (with binary parameters shown in the
title). If we discard all times below the point where the
orbit-averaged frequency $\langle
f_{22}\rangle\equiv\avgOmega/(2\pi)$ (pink dashed curve) crosses
$\fLow=20$ Hz, only the pink section is retained and the blue section
is discarded even though it contains some frequencies above $20$
Hz. On the other hand, using
$f_{22}^{\text{p}}\equiv\omegaP/(2\pi)$ (blue dashed curve) to
pick this time ensures that the discarded region (orange) contains no
frequencies above $20$ Hz.
}
\label{fig:setting_flow}
\end{figure}

For quasicircular waveform models with only the $\dm$ mode, $\tLow$ can be
chosen to be the time when
\begin{equation}
\label{eq:omega22_flow}
\omegatwotwo(\tLow) = 2 \pi \, \fLow.
\end{equation}
Because $\omegatwotwo(t)$ is a monotonically increasing function for
quasicircular binaries, frequencies $>\fLow$ only occur at times $>\tLow$. This
is no longer the case for eccentric binaries as $\omegatwotwo(t)$ can be
nonmonotonic. An example is shown in Fig.~\ref{fig:setting_flow}, where we see
that $\omegatwotwo(t)/(2\pi)$ crosses $\fLow=20$ Hz at several different times.
One could choose the earliest of these crossings as $\tLow$, but this only works
if the original waveform is long enough to include all such crossings. If the
original waveform only includes a subset of the crossings, this approach cannot
guarantee that the discarded waveform only contains frequencies $<\fLow$.  To
ensure all frequencies above $\fLow$ are included, we need to generalize
Eq.~\eqref{eq:omega22_flow} to eccentric binaries.

A seemingly natural choice is to replace $\omegatwotwo(t)$ in
Eq.~\eqref{eq:omega22_flow} with the monotonically increasing $\avgOmega(t)$
from Eq.~\eqref{eq:mean_motion}:
\begin{equation}
\label{eq:setting_flow_omega_avg}
  \avgOmega(\tLow) = 2\pi \, \fLow,
\end{equation}
The pink dashed line in Fig.~\ref{fig:setting_flow} shows $\avgOmega/(2\pi)$,
and the frequencies retained when setting $\tLow$ using
Eq.~\eqref{eq:setting_flow_omega_avg} are also marked in pink. However, in this
approach the section colored in blue is discarded, even though it still
includes some frequencies above $\fLow=20$ Hz.

Instead, we propose that $\tLow$ should be set using the interpolant through
pericenter frequencies, $\omegaP(t)$, which is already constructed when
evaluating Eqs.~\eqref{eq:e_omegatwotwo} and \eqref{eq:eGW}.
\begin{equation}
\label{eq:setting_flow_omega_peri}
    \omegaP(\tLow) = 2\pi \, \fLow.
\end{equation}
Because $\omegaP(t)$ represents the upper envelope of $\omegatwotwo(t)$, this
approach guarantees that the discarded waveform ($t<\tLow$) does not contain
any frequencies $>\fLow$. This is demonstrated in Fig.~\ref{fig:setting_flow},
where we see that the blue section is included if
Eq.~\eqref{eq:setting_flow_omega_peri} is used to set $\tLow$.

So far, we only considered the $\dm$ mode when determining $\tLow$. The
frequency of the $(\ell, m)$ waveform mode (Eq.~\eqref{eq:h}) can be
approximated during the inspiral as $\omega_{\ell m}(t) \sim (m/2) ~
\omegatwotwo(t)$~\cite{Blanchet:2013haa}.
Therefore, for models containing higher modes,
Eq.~\eqref{eq:setting_flow_omega_peri} should be replaced with:
\begin{equation}
\label{eq:setting_flow_omega_peri_hm}
    \omegaP(\tLow) = \left(\frac{2}{m_{\text{max}}}\right) \, 2\pi \, \fLow,
\end{equation}
where $m_{\text{max}}$ is the largest $m$ among all included modes.

\subsection{Summary}
Our procedure to compute the eccentricity and mean anomaly from the waveform
can be summarized as follows:
\begin{enumerate}
\item Find the times corresponding to the pericenters and apocenters, which we
    denote as $\{\tP_{i}\}$ and $\{\tA_{i}\}$, respectively. In the example in
    Fig.~\ref{fig:ecc_definition}, $\{\tP_{i}\}$ and $\{\tA_{i}\}$ are
    identified as the local maxima and minima, respectively, of $\omegatwotwo$,
    but other methods for locating these times will be discussed in
    Sec.~\ref{sec:methods_to_locate_extrema}.
\item Evaluate $\omegatwotwo(t)$ at $\{\tP_{i}\}$ and $\{\tA_{i}\}$ to
  get the frequencies at pericenters and apocenters and construct
  interpolants in time, $\omegaP(t)$ and $\omegaA(t)$, using these
  data. We use cubic splines for interpolation.\footnote{When the
      number of pericenters or apocenters in not sufficient to build a cubic
      spline, the order of the spline is reduced accordingly.}
\item Obtain $\etwotwo(t)$ using $\omegaP(t)$ and $\omegaA(t)$ in
    Eq.~(\ref{eq:e_omegatwotwo}). Finally, apply the transformation in
    Eq.~\eqref{eq:eGW} to obtain the eccentricity $\egw(t)$.
\item Use the pericenter times $\{\tP_{i}\}$ in
    Eq.~\eqref{eq:mean_anomaly_definition} to compute the mean anomaly
    $\lgw(t)$.
\item To get the eccentricity and mean anomaly at a reference frequency
    $\fref$, first use the orbit averaged frequency $\avgOmega(t)$
    (Eq.~\eqref{eq:mean_motion}) to get the corresponding $\tref$. However,
    instead of using a fixed $\fref$ in Hz, a fixed dimensionless frequency or
    time, or a fixed number of orbits before a dimensionless frequency/time
    might be a better choice for eccentric binaries
    (Sec.~\ref{sec:selecting_reference_point}).
\item Use $\omegaP(t)$ (Eq.~\eqref{eq:setting_flow_omega_peri_hm}) to truncate
    time-domain signals at a given start frequency $\fLow$ so that the
    discarded waveform does not contain any frequencies above $\fLow$.
\end{enumerate}

%==========================================================================
\section{Methods to locate pericenters and apocenters}
\label{sec:methods_to_locate_extrema}

In Sec.~\ref{sec:defining_eccentricity} and Fig.~\ref{fig:ecc_definition}, the
pericenter and apocenter times are taken to correspond to local extrema in
$\omegatwotwo(t)$. Identifying these times is a crucial step in our definitions
of eccentricity and mean anomaly, as well as the generalizations of $\fref$ and
$\fLow$. In this section, we explore several different alternatives for
identifying the pericenter/apocenter times and their benefits and drawbacks.
Instead of $\omegatwotwo(t)$, these methods set extrema in various other
waveform quantities (like the amplitude) as the pericenter/apocenter times.
Therefore, the pericenter/apocenter times can depend on the method used, and
each of these alternatives should be viewed as a new \emph{definition} of
eccentricity and mean anomaly. However, all of these methods satisfy the
criteria listed in Sec.~\ref{sec:defining_eccentricity} for a good definition
of eccentricity, and as we will show in Sec.~\ref{sec:tests} the differences
between the different methods are generally small.  We denote the waveform
quantity whose extrema are used as $\uu(t)$. Given $\uu(t)$, we use the
\texttt{find\_peaks} routine within \texttt{SciPy}~\cite{2020SciPy-NMeth} to
locate the extrema.

\subsection{Frequency and amplitude}
\label{sec:frequency_and_amplitude}

The most straightforward choice for $\uu(t)$ is
\begin{equation}
    \uu(t) = \omegatwotwo(t),
\end{equation}
as considered in Fig.~\ref{fig:ecc_definition}. The local maxima in $\uu(t)$ are
identified as the pericenters while the local minima are identified as
apocenters. We refer to this method as the \mFreq{} method.

Because $\omegatwotwo(t)$ relies on a time derivative -- see
Eq.~\eqref{eq:omega_22} -- it can be noisy in some cases, especially for NR
waveforms. Such noise can lead to spurious extrema in $\omegatwotwo(t)$ that
can be mistaken for pericenters/apocenters. Such problems can be avoided by
locating the extrema of the amplitude of the $\dm$ mode, i.e.
\begin{equation}
    \uu(t) = \Atwotwo(t).
\end{equation}
We refer to this method as the \mAmp{} method and recommended it over the
\mFreq{} method.

\begin{figure}
  \centering
  \includegraphics[width=\columnwidth,trim=0 8 0 2]{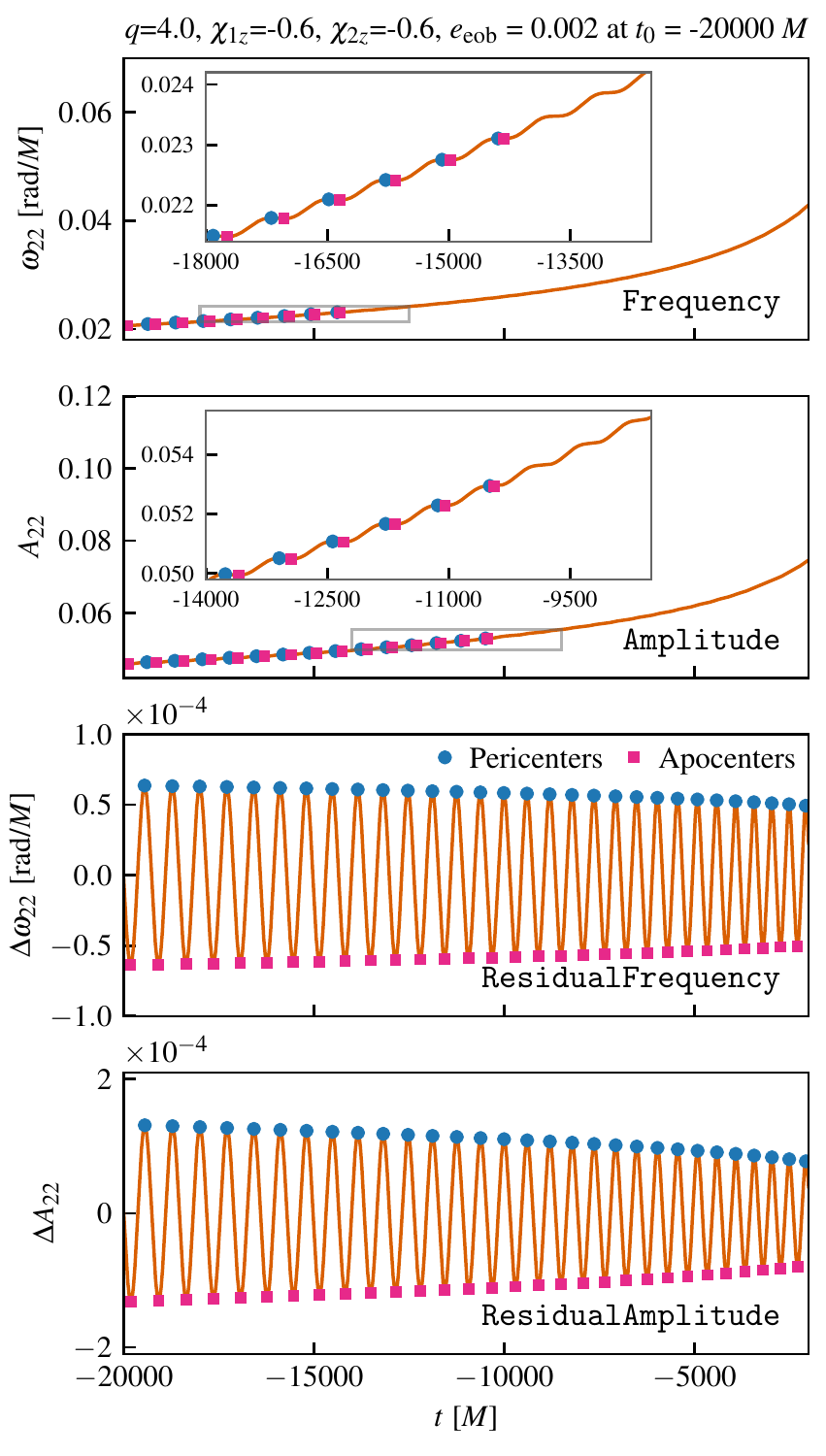}
\caption{
Limitations of the \mAmp{} and \mFreq{} methods in identifying pericenters
(\cPeri{} \markPeri{}s) and apocenters (\cApo{} \markApo{}s) for a low eccentricity
waveform. These methods (top two rows) detect only the first few
pericenters/apocenters and fail once sufficient eccentricity is radiated away.
On the other hand, the \mResAmp{} and \mResFreq{} methods (bottom two rows) can
detect all of the pericenters/apocenters present. The waveform is generated
using \SEOB{} and the binary parameters are given in the title.
}
\label{fig:amp_vs_res_amp}
\end{figure}

The simplicity of the \mFreq{} and \mAmp{} methods comes with the drawback that
these methods fail for small eccentricities, as illustrated in
Fig.~\ref{fig:amp_vs_res_amp}. The top two rows show $\omegatwotwo$ and
$\Atwotwo$ for an eccentric \SEOB{}~\cite{Ramos-Buades:2021adz} waveform. While
local extrema can be found at early times, as eccentricity is radiated away,
the prominence of the extrema decreases until local extrema cease to exist. The
onset of this breakdown is signaled by the pericenters and apocenters
converging towards each other, as seen in the figure insets. This occurs
because at small eccentricity, the secular growth in $\omegatwotwo$ and
$\Atwotwo$ dominates the modulations due to eccentricity. We find that for
eccentricities $\egw \lesssim 10^{-2} \ldots 10^{-3}$ (see Sec.~\ref{sec:tests}),
the \mFreq{} and \mAmp{} methods can fail to measure the eccentricity. This
breakdown point can be approximately predicted by the following
order-of-magnitude estimate.

\subsubsection{Estimating the breakdown point of the \mFreq{} method}

The inspiral rate of a binary in quasicircular orbit at Newtonian order
is given by (e.g.\ \cite{Blanchet:2013haa})
% Start with Blanchet Eq. (318) [Phi=-1/nu Omega^{5/8}]
% take two time-derivatives, substitute in Theta^{-1/4}=4x and use Eq. (226)
\begin{equation}\label{eq:omega22-inspiral}
\frac{\dd \omega_{22}^{\rm circ}}{\dd t} = \frac{192}{5}\nu\frac{1}{M^2} \left(\frac{M\zeroOmega}{2}\right)^{11/3},
\end{equation}
where $\nu=q/(1+q)^2$ is the symmetric mass ratio.

For small eccentricities, eccentricity induces an oscillatory component to the
frequency,
\begin{equation}\label{eq:delta_omega}
\omegatwotwo(t) \approx\zeroOmega(t)+ A \sin(\omega_r t),
\end{equation}
where $\omega_r$ denotes the radial oscillation frequency.  The amplitude $A$
of the oscillations can be related to eccentricity by substituting into
Eq.~(\ref{eq:e_omegatwotwo}) and expanding to first order in $A$, yielding
$A=2\etwotwo \, \zeroOmega$.  For a given short time interval, we
take $A$ to be constant.

Extrema in $\omegatwotwo(t)$ correspond to zeros of the time derivative
\begin{equation}\label{eq:dt_eccentric_omega22}
\frac{\dd\omegatwotwo}{\dd t} \approx\frac{\dd\omega_{22}^{\rm circ}}{\dd t}+ A\omega_r \cos(\omega_r t).
\end{equation}
Such zeros exist only if the oscillatory component dominates over the inspiral part, $A\omega_r\gtrsim \dd \omega_{22}^{\rm circ}/\dd t$, i.e.\ for sufficiently large eccentricities:
\begin{equation}
\etwotwo\gtrsim \frac{48}{5}\nu \left(\frac{M\omegatwotwo}{2}\right)^{5/3}\frac{\omegatwotwo}{2\omega_r}.
\end{equation}
%% \begin{equation}\label{eq:delta_omega}
%% \delta\omegatwotwo \approx  2 \, \etwotwo \, \omegatwotwo \, \sin(\omega_r t),
%% \end{equation}
%% where $\omega_r$ denotes the radial oscillation frequency and we have
%% related the amplitude of the oscillation to $\etwotwo$ using
%% Eq.~(\ref{eq:e_omegatwotwo}) \vv{Not sure I follow...dumb it down further?}.  Equation~(\ref{eq:delta_omega}) implies
%% a maximum time-derivative of the oscillatory component of
%% \begin{equation}\label{eq:max-delta-omega22}
%% \max\left|\frac{d\delta\omegatwotwo}{dt}\right|\approx 2 \, \etwotwo \, \omegatwotwo \, \omega_r.
%% \end{equation}
%% In order to have local extrema in $U(t)=\omegatwotwo(t)$, the oscillatory
%% component must dominate the secular inspiral rate, i.e. the left-hand side of
%% Eq.~(\ref{eq:max-delta-omega22}) must be larger than the left hand side of
%% Eq.~(\ref{eq:omega22-inspiral}). This implies
%% \begin{equation}
%% \etwotwo\gtrsim \frac{48}{5}\nu \left(\frac{M\omegatwotwo}{2}\right)^{5/3}\frac{\omegatwotwo}{2\omega_r}.
%% \end{equation}
Here we have dropped the subscript ``circ'', as $\zeroOmega\approx
\omegatwotwo$ at leading order in the assumed small eccentricity.  Neglecting
pericenter advance, i.e.\ setting $\omegatwotwo/(2\omega_r)=1$, and noting that
for small eccentricity, $\etwotwo\approx (3/4) \, \egw$ (Eq.~\ref{eq:eqE3}), we
find that local extrema in $\omegatwotwo(t)$ are only present if
\begin{equation}\label{eq:e-breakdown}
\egw\gtrsim \frac{192}{15}\nu\left(\frac{M\omegatwotwo}{2}\right)^{5/3}.
\end{equation}
The systems considered in this paper have $\omegatwotwo\sim
0.02/M\ldots 0.1/M$ (e.g.\ Figs.~\ref{fig:ecc_definition}
or \ref{fig:amp_vs_res_amp}), so that for comparable mass binaries,
Eq.~(\ref{eq:e-breakdown}) predicts a breakdown of the \mFreq{}
method for $\egw \sim 10^{-3}\ldots 10^{-2}$.

This motivates us to consider alternative methods to detect local extrema that
also work for small eccentricities. In the following, we will consider
different methods that first subtract the secular growth in $\omegatwotwo$ or
$\Atwotwo$, and use the remainder as $\uu(t)$.

\subsection{Residual frequency and residual amplitude}
\label{sec:residual_frequency_and_residual_amplitude}

We begin with a simple extension of the \mFreq{} method, which we refer to as
the \mResFreq{} method:
\begin{equation}
\label{eq:residual_omega22}
  \uu(t) = \resOmega(t) \equiv \omegatwotwo(t) - \zeroOmega(t),
\end{equation}
and likewise the \mResAmp{} method:
\begin{equation}
\label{eq:residual_amplitude}
    \uu(t) = \resAmp(t) \equiv \Atwotwo(t) - \zeroA(t),
\end{equation}
where $\zeroOmega$ and $\zeroA$ are the frequency and amplitude of the $\dm$
mode for a quasicircular counterpart of the eccentric binary. We define the
quasicircular counterpart as a binary with the same component masses and spins,
but with zero eccentricity. The time array of the quasicircular waveform is
shifted so that its peak time coincides with that of the eccentric waveform.
Once again, the local maxima in $\uu(t)$ are identified as the pericenters while
the local minima are identified as apocenters.

\begin{figure}
\centering
\includegraphics[width=\columnwidth,trim=0 5 0 2]{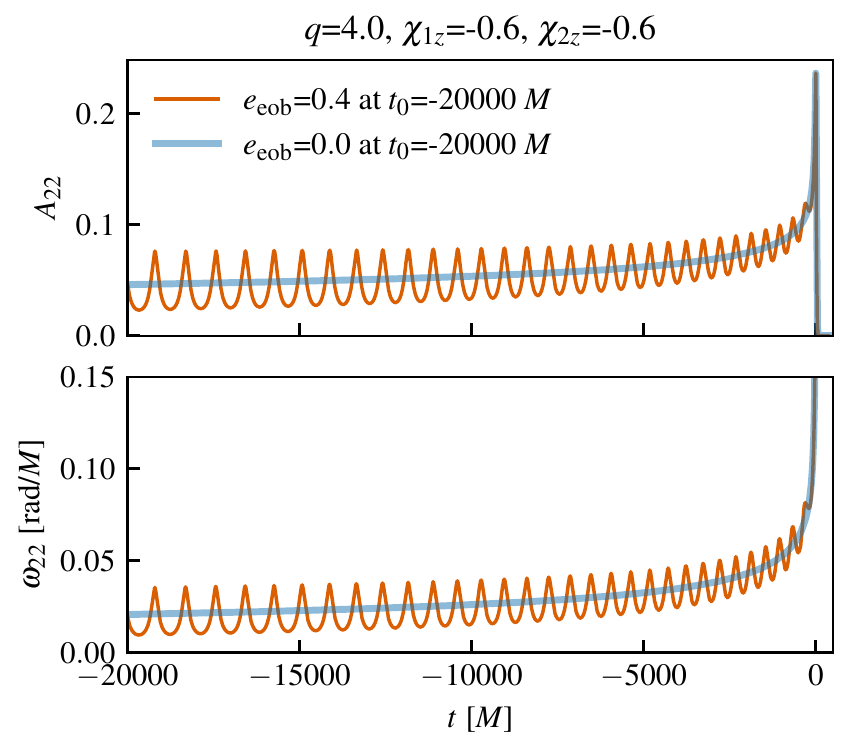}
\caption{
Comparison of the amplitude (top) and the frequency (bottom) of an eccentric
\SEOB{} waveform to those of its quasicircular counterpart. The binary
parameters are shown in the figure text. Both waveforms are aligned so that
$t=0$ occurs at the peak of $\Atwotwo$. The quasicircular counterpart captures
the secular growth in the amplitude and frequency of the eccentric waveform.
}
\label{fig:eccentric_and_non_eccentric}
\end{figure}

Eqs.~\eqref{eq:residual_omega22} and \eqref{eq:residual_amplitude} are
motivated by the observation~\cite{Islam:2021mha} that the quasicircular
counterpart waveform captures the secular trend of the eccentric waveform, when
the peak times of the waveforms are aligned. This is demonstrated for an
example eccentric \SEOB{} waveform in
Fig.~\ref{fig:eccentric_and_non_eccentric}. The quasicircular counterpart falls
approximately at the midpoint between the peaks and troughs of amplitude and
frequency of the eccentric waveform. We find this to be the case for the full
range of eccentricities, and waveforms of all origins.

For an eccentric waveform model, the quasicircular counterpart can be easily
generated by evaluating the model with eccentricity set to zero while keeping
the other parameters fixed. For eccentric NR waveforms, such a
quasicircular NR waveform may not exist and one can use a quasicircular
waveform modelto generate the quasicircular counterpart. In
this paper, we use the \PhenomT{}~\cite{Estelles:2021gvs} quasicircular
waveform model to generate quasicircular counterparts of NR waveforms and
\PhenomT{} is currently set as the default choice in \package{} as it supports
a wide range of values for the binary parameters. One can also use
more accurate models like the NR surrogate model
\texttt{NRHybSur3dq8}~\cite{Varma:2018mmi} whenever the parameters fall within
the regime of validity of the surrogate model. Similarly to how the different
methods to locate extrema are part of the eccentricity definition, the choice
of quasicircular model should also be considered to be a part of the
definition. The impact of the choice of the quasicircular model on eccentricity
is generally small and will be explored further in
Sec.~\ref{sec:robustness_dependence_on_methods}.

By first subtracting the secular growth in the eccentric waveform, the
\mResFreq{} and \mResAmp{} methods can detect local extrema even for
small eccentricities. The bottom two rows of Fig.~\ref{fig:amp_vs_res_amp} show
an example where these methods succeed while the \mFreq{} and \mAmp{} methods
fail. Once again, between \mResFreq{} and \mResAmp{}, we recommend \mResAmp{}
as it is less prone to numerical noise for NR waveforms. While the \mResFreq{}
and \mResAmp{} are robust and straightforward to implement, their main drawback
is that they require the evaluation of a quasicircular waveform, which
increases the computational expense. We consider the next set of methods to model
the secular trend without relying on additional waveform evaluations.

\begin{figure*}[htb!]
  \centering
  \includegraphics[width=0.96\textwidth,trim=0 5 0 4]{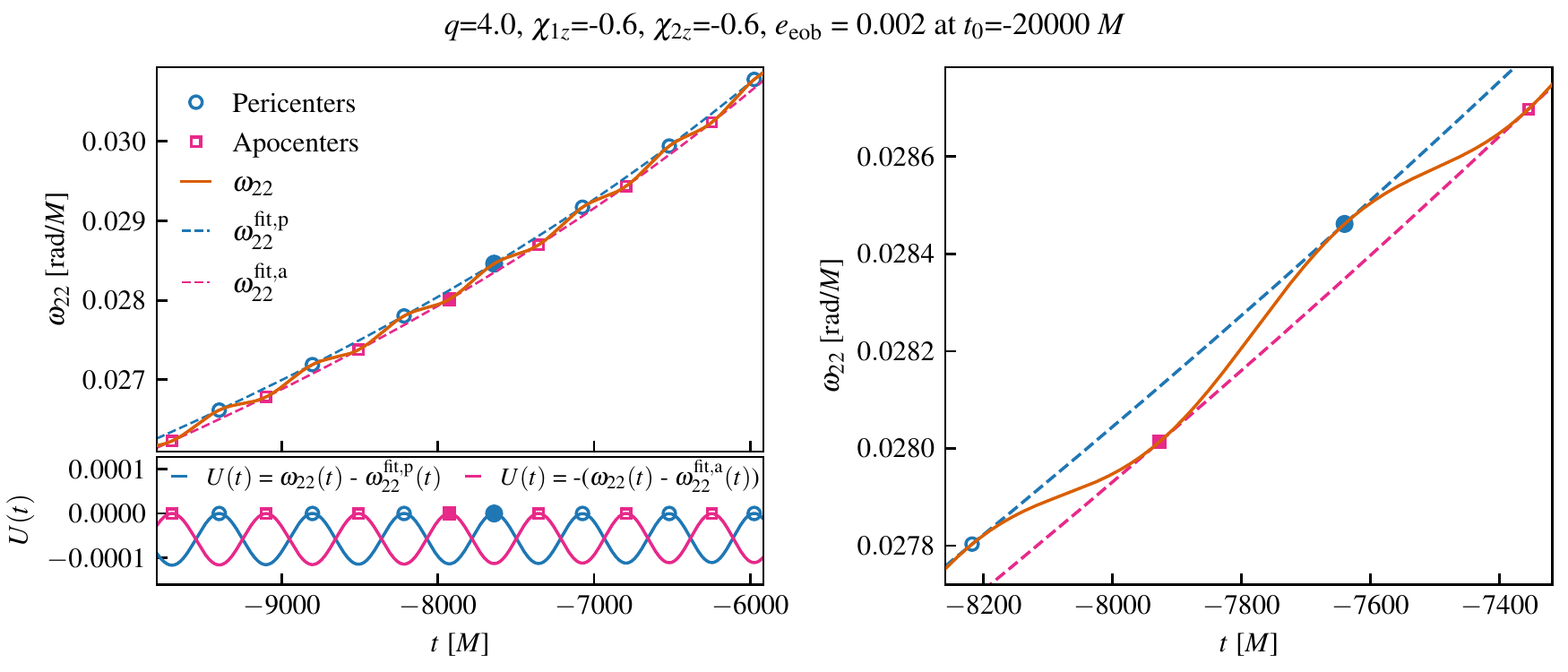}
\caption{Illustration of the \mFreqFits{} method. {\itshape Left:} The
\cPeri{} circles indicate the $2N+1=7$ extrema through which the
fitting function Eq.~(\ref{eq:fit}) passes.  The lower panel shows the
envelope-subtracted data from which the extrema $T_\alpha$ are
determined.  The solid \cPeri{} \markPeri{} indicates the central extremum,
whose parameters are used for the eccentricity definition.  The
\cApo{} \markApo{} and the \cApo{} dashed line show the analogous
construction for the apocenter passages. {\itshape Right:} Enlargement of
the region around the solid markers in the upper panel on the
left. The waveform is generated using \SEOB{}, and the binary parameters
are given in the title.
}
\label{fig:frequencyfits}
\end{figure*}

\subsection{Frequency fits and amplitude fits}\label{sec:frequenc-fits}

The \mResAmp{} and \mResFreq{} methods described in
Sec.~\ref{sec:residual_frequency_and_residual_amplitude} have the disadvantage
that they require a quasicircular reference waveform for subtraction. Such a
reference waveform may not be available, or deviations in the reference
waveform may lead to differences in the recovered eccentricity (see
Sec.~\ref{sec:robustness_dependence_on_methods}).

The \mFreqFits{} method avoids the need for a reference waveform by
self-consistently fitting the envelopes $\omegaP(t)$ (for pericenters) and
$\omegaA(t)$ (for apocenters) that appear in
Fig.~\ref{fig:ecc_definition}, an idea introduced in  Lewis
\etal{}~\cite{Lewis:2016lgx}. To simplify the explanation, we will first
describe this method when applied to locate pericenters. The idea lies in
considering a \textit{local} stretch of data $\omegatwotwo(t)$ for $t\in[t_L,
t_R]$, 
in which we identify the times $\eccfitTp_\alpha$ (labeled by $\alpha$) as
local maxima of the envelope-subtracted frequency
(Eq.~(\ref{eq:envelope-subtracted})), while self-consistently constructing the
envelope fit $\omegaFitP(t)$ through $\omegatwotwo(t)$ evaluated at
$\eccfitTp_\alpha$. The fit $\omegaFitP(t)$, the local maxima times
$\eccfitTp_\alpha$, and the interval $[t_L, t_R]$ are iteratively refined and
the central $\eccfitTp_\alpha$ is identified as a pericenter time.
%containing pericenter times $\eccfitTp_\alpha$ (labeled
%by $\alpha$), which in turn are fitted by $\omegaFitP(t)$.  The deviation from
%this fit --Eq.~(\ref{eq:envelope-subtracted}) below-- defines the
%$\eccfitTp_\alpha$, so that these elements must be determined
%self-consistently.  Another element of self-consistency is introduced through
%the interval $[t_L, t_R]$, which is adjusted to encompass a certain number of
%maxima $\eccfitTp_\alpha$.
%, and the time-interval $[t_L, t_R]$ is chosen to contain a fixed number of maxima $\eccfitTp_\alpha$.

To make this idea precise, we start by choosing a time  $\hatT$, which will
roughly correspond to the middle of the fitting interval.  We now seek to
determine a fitting function $\omegaFitP(t)$ through the pericenter
frequencies, valid in a time-interval $[t_L, t_R]$ encompassing $\hatT$, as
well as times $\eccfitTp_\alpha\in[t_L, t_R]$, $\alpha=0, \ldots, 2N$ (with
$N=3$, as explained after Eq.~(\ref{eq:reparameterization1})).  These
quantities are determined in a self-consistent manner such that the following
conditions are all satisfied:
\begin{enumerate}
    \item  \label{cond:1} $\eccfitTp_\alpha$ are local maxima of the envelope-subtracted frequency $\uu(t)$ given by:
  \begin{equation}
    \label{eq:envelope-subtracted}
    \uu(t) = \omegatwotwo(t) - \omegaFitP(t).
  \end{equation}

\item\label{cond:2}  $\omegaFitP(t)$ is a fit through the $2N+1$ evaluations of $\omegatwotwo(t)$ at
    times $\eccfitTp_\alpha$, i.e. $(\eccfitTp_\alpha, \omegatwotwo(\eccfitTp_{\alpha}))$
   in the interval $[t_L, t_R]$,
  \begin{equation}
    \label{eq:omega-fit}
    \omegaFitP(\eccfitTp_\alpha) \approx \omegatwotwo(\eccfitTp_{\alpha}),\quad \alpha=0, \ldots, 2N.
\end{equation}
\item\label{cond:3} The time-interval $[t_L, t_R]$ contains precisely $2N+1$
    local maxima of $U(t)$ where the first $N$ are before $\hatT$, and
    the others after.
\end{enumerate}

If these conditions are met, then the extremum in the middle, $(\eccfitTp_N,
\omegatwotwo(\eccfitTp_N))$ will be identified as a pericenter passage, and
included in the overall list of pericenters for the inspiral.

This procedure is illustrated in Fig.~\ref{fig:frequencyfits}.  The top panel
shows $\omegatwotwo(t)$ in orange, for a configuration with
eccentricity so small that $\omegatwotwo(t)$ does not have extrema.  The
locations of the identified local maxima $\big(T_\alpha$,
$\omegatwotwo(T_\alpha)\big)$ are indicated by \cPeri{} circles, with the middle one
(corresponding to $T_N$) being filled. The lower panel shows the envelope
subtracted function, whose maxima determine the $T_\alpha$.

In practice, the fitting function is chosen to have the functional form
\begin{equation}\label{eq:fit}
\omegaFitP(t; \,A,\,n,\,\eccfitT) = A(\eccfitT-t)^n,
\end{equation}
with fit-parameters $\{A, n, \eccfitT\}$.  The form of Eq.~(\ref{eq:fit}) is
inspired by the leading order PN behavior of a
quasicircular binary inspiral, which has the form of Eq.~(\ref{eq:fit}) with
exponent $-3/8$~\cite{Blanchet:2013haa}.  In addition, Eq.~(\ref{eq:fit})
ensures monotonicity by construction.  To reduce correlations between the
parameters $A$ and $n$, the fitting function is reparameterized by $\{f_0, f_1,
\eccfitT\}$ where $f_0$ and $f_1$ represent the function value and first
time-derivative at a time $\eccfitTmid$,
\begin{subequations}\label{eq:reparameterization1}
\begin{align}
  f_0=& A(\eccfitT-\eccfitTmid)^n,\\
  f_1=& -n A(\eccfitT-\eccfitTmid)^{n-1} = -n \frac{f_0}{\eccfitT-\eccfitTmid}.
\end{align}
\end{subequations}
Equations~(\ref{eq:reparameterization1}) are readily inverted to yield
\begin{subequations}
\begin{align}
n=& -\frac{f_1(\eccfitT-\eccfitTmid)}{f_0},\\
A=& f_0(\eccfitT-\eccfitTmid)^{-n}.
\end{align}
\end{subequations}
The fit for $\{f_0, f_1, \eccfitT\}$ is performed with the \texttt{curve\_fit}
routine of the \texttt{SciPy}~\cite{2020SciPy-NMeth} library.
Because there are three free parameters, at least three local maxima are
needed to perform the fit; we choose $2N+1=7$ maxima for increased robustness.
The concrete choice for $\eccfitTmid$ is found to be not critical; we choose
the time in the middle of the entire waveform to be analyzed.

To analyze an entire waveform, we proceed from the start of the waveform toward
the merger.  At the first, ``cold'' initialization at the start of the waveform, we
choose $t_L$ to be the start of the waveform, $\hatT$ to be $N$ orbits later
(as judged by the accumulated $\phitwotwo$), and $t_R$ to be $2N$ orbits later.
We initialize a first guess for $\omegaFitP$ through a fit to 
$\omegatwotwo(t)$ during the first 10 orbits of the waveform.

In order to satisfy the conditions \ref{cond:1} to \ref{cond:3}
self-consistently, an iterative procedure is applied: local maxima of
$U(t)$ are calculated using \texttt{find\_peaks}, and the interval
$[t_L,t_R]$ is adjusted to achieve the desired number of extrema on
either side of $\hatT$.\footnote{For the very first application of
this procedure at the start of the waveform, $t_L$ cannot be reduced
to before the start of the waveform, so if needed we increase $\hat t$
instead.}  Now an improved $\omegaFitP$ is computed by fitting to the
extrema, Eq.~(\ref{eq:omega-fit}), and the procedure is iterated until
the changes in the extrema $\eccfitTp_\alpha$ and fitting parameters
$\{f_0,f_1,T\}$ fall below a tolerance, typically $10^{-8}$.  At the
initial cold start, this typically takes 3-5 iterations. 

We then shift the analysed region by one pericenter passage at a time,
i.e.\ $\hatT\to \hatT=(\eccfitTp_N+\eccfitTp_{N+1})/2$, $t_L\to
(\eccfitTp_0 + \eccfitTp_1)/2$, $t_R\to \eccfitTp_{2N}+
1.5\,\into\,(\eccfitTp_{2N}-\eccfitTp_{0})/(2N)$, and repeat the
iterative procedure to satisfy conditions \ref{cond:1}
to \ref{cond:3}, using the current $\omegaFitP$ as the initial guess.
Because of the improved guess for $\omegaFitP$, each successive
pericenter passage needs only 2--3 iterations to converge.  We stop the
procedure when $t_L$ reaches the end of the waveform, or when all three
conditions can no longer be simultaneously satisfied.  For instance, in rare cases,  the
iterative procedure settles into a limiting cycle, which switches between two
different results for the interval  $[t_L, t_R]$, the extrema $T_\alpha$, and
the fit $\omegaFitP$.

%% \TODO{OLD VERSION BELOW}
%% In order to satisfy the conditions \ref{cond:1} to \ref{cond:3}
%% self-consistently, a iterative procedure is applied: Starting with an
%% initial fit $\omegaFitP$ \vv{How do you do the initial fit if you have no pericenters to start with?}, local maxima of $U(t)$ are calculated
%% using \texttt{find\_peaks}, and the interval $[t_L,t_R]$ is adjusted
%% to achieve the desired number of extrema.  Now an improved fit
%% $\omegaFitP$ is performed, and the procedure iterated until all
%% quantities \vv{[List out these quantities?]} remain unchanged.  This typically takes 2-3 iterations.

%% To trace out a full inspiral, we begin by applying this fitting
%% procedure at the start of the waveform.  In this initial step, we
%% initialize $\omegaFitP$ to a fit to the full $\omegatwotwo(t)$, and we
%% increase $\hatT$ so that it lies after the first $N$ local maximum, to
%% satisfy condition \ref{cond:3}.  We then shift $\hatT$ by one maximum
%% at a time, i.e. $\hatT\to \hatT=(\eccfitTp_N+\eccfitTp_{N+1})/2$
%% \TODO{The replacement $T_\alpha$ by $\eccfitTp_\alpha$ leads to weird ambiguities here}
%% and reapply the
%% iterative fitting procedure.  This process continues until
%% all three conditions \ref{cond:1}--\ref{cond:3} can no longer be satisfied.
%% This happens near the merger when there simply are no longer enough
%% maxima after $\hatT$ to satisfy condition \ref{cond:3}.

Equation~(\ref{eq:envelope-subtracted}) identifies local maxima of
$\omegatwotwo(t) - \omegaFitP(t)$, i.e.\ pericenter passages.  To identify apocenter
passages, we \textit{change the sign} of the right-hand-side of
Eq.~(\ref{eq:envelope-subtracted}), while keeping the remainder of the
algorithm unchanged.  The algorithm will then generate a fit to the
apocenter points, $\omegaFitA$, as indicated in \cApo{} in
Fig.~\ref{fig:frequencyfits}.

The procedure outlined above also works if we fit the amplitude $\Atwotwo$
in place of $\omegatwotwo$, since at leading post-Newtonian order, the
amplitude also has the form of Eq.~(\ref{eq:fit}) with exponent
$-1/4$~\cite{Blanchet:2013haa}. We refer to the method of finding the
pericenters/apocenters by fitting to $\Atwotwo$ as \mAmpFits{}. Once again,
\mFreqFits{} is more prone to numerical noise as it relies on $\omegatwotwo$.
Therefore, we recommend \mAmpFits{} over \mFreqFits{}.

%==========================================================================
\section{Robustness tests}
\label{sec:tests}

In this section, we check the robustness of our eccentricity definition and the
different methods to locate pericenters/apocenters by putting our
implementation through various tests.

\subsection{The large mass ratio limit of $\egw$}
\label{sec:egw_emri_limit}

In Sec.~\ref{sec:introduction}, we noted that one of the desired but not
strictly required features of an ideal eccentricity definition is that in the
limit of large mass ratio, it should approach the test particle eccentricity on
a Kerr geodesic. The geodesic eccentricity $\egeo$ typically used for EMRI
calculations~\cite{1959RSPSA.249..180D, 1961RSPSA.263...39D} is given by:
\begin{equation}
  \egeo = \frac{r^{\text{a}} - r^{\text{p}}}{r^{\text{a}} + r^{\text{p}}},
\end{equation}
where $r^{\text{p}}$ and $r^{\text{a}}$ are the pericenter and apocenter
separations along the geodesic in Boyer--Lindquist coordinates. To test the
test particle limit of $\egw$, we compare $\egw$ and $\egeo$ for an EMRI
waveform with $q=\infty$ and nonspinning BHs, but with varying eccentricities
in the range $\egeo \in [0, 0.5]$. In the $q\to\infty$ limit, there is no
orbital evolution and the waveform is that of a test particle following a
geodesic. For our comparisons, we use the waveforms computed within this
framework in Ref.~\cite{Ramos-Buades:2022lgf} using a frequency domain
Teukolsky code. Because there is no orbital evolution these waveforms each have
a constant value of eccentricity $\egeo$ and orbit averaged frequency
$\avgOmega$.

\begin{figure}%[thb]
  \centering
  \includegraphics[width=\columnwidth]{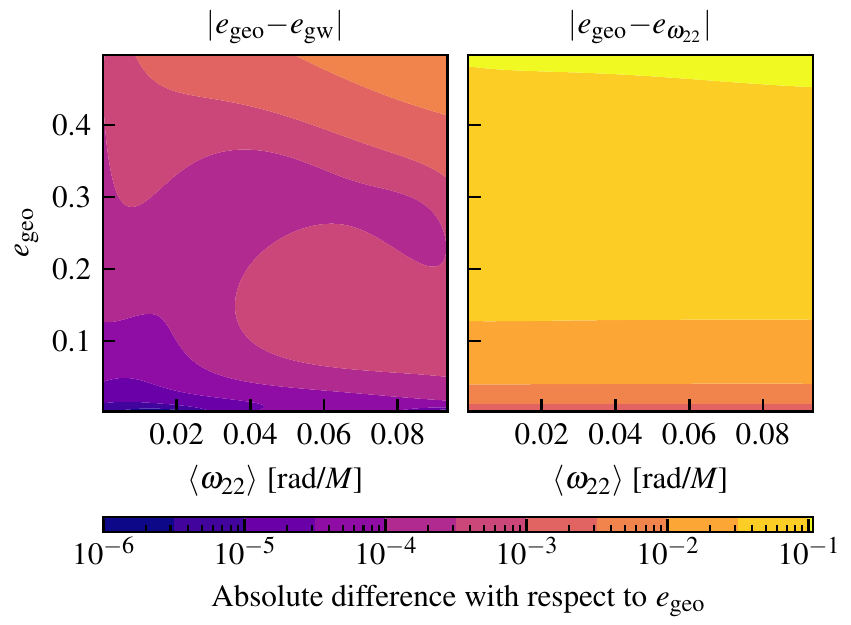}
\caption{
Comparison of $\egw$ and $\etwotwo$ to the geodesic eccentricity $\egeo$ in the
$q\to\infty$ limit, as a function of the orbit averaged frequency $\avgOmega$.
In the left panel, the colors show the absolute difference between $\egeo$ and
$\egw$ measured using Eq.~(\ref{eq:eGW}) with the \mAmp{} method. The right
panel shows the same for $\etwotwo$. $\egeo$ is closer to $\egw$ than
$\etwotwo$ by about two orders of magnitude.
}
\label{fig:e_gw_vs_e_emri}
\end{figure}

Figure~\ref{fig:e_gw_vs_e_emri} shows the differences $|\egeo - \egw|$ and
$|\egeo - \etwotwo|$, evaluated at different values of $\egeo$ and
$\avgOmega$. While $\egw$ does not exactly match $\egeo$ in the test particle
limit, the differences for $\egw$ lie in the range $\sim[10^{-6},
6\into10^{-3}]$, whereas the differences for $\etwotwo$ lie in the range
$\sim[5\into10^{-4}, 10^{-1}]$. Therefore, $\egw$ is an improvement over
$\etwotwo$ in two ways: $\egw$ has the correct Newtonian limit (as shown by
Ref.~\cite{Ramos-Buades:2021adz}) and is closer to $\egeo$ in the test particle
limit, by about two orders of magnitude.

\begin{figure*}[htb]
\centering
\includegraphics[width=0.95\textwidth,trim=0 3 0 4]{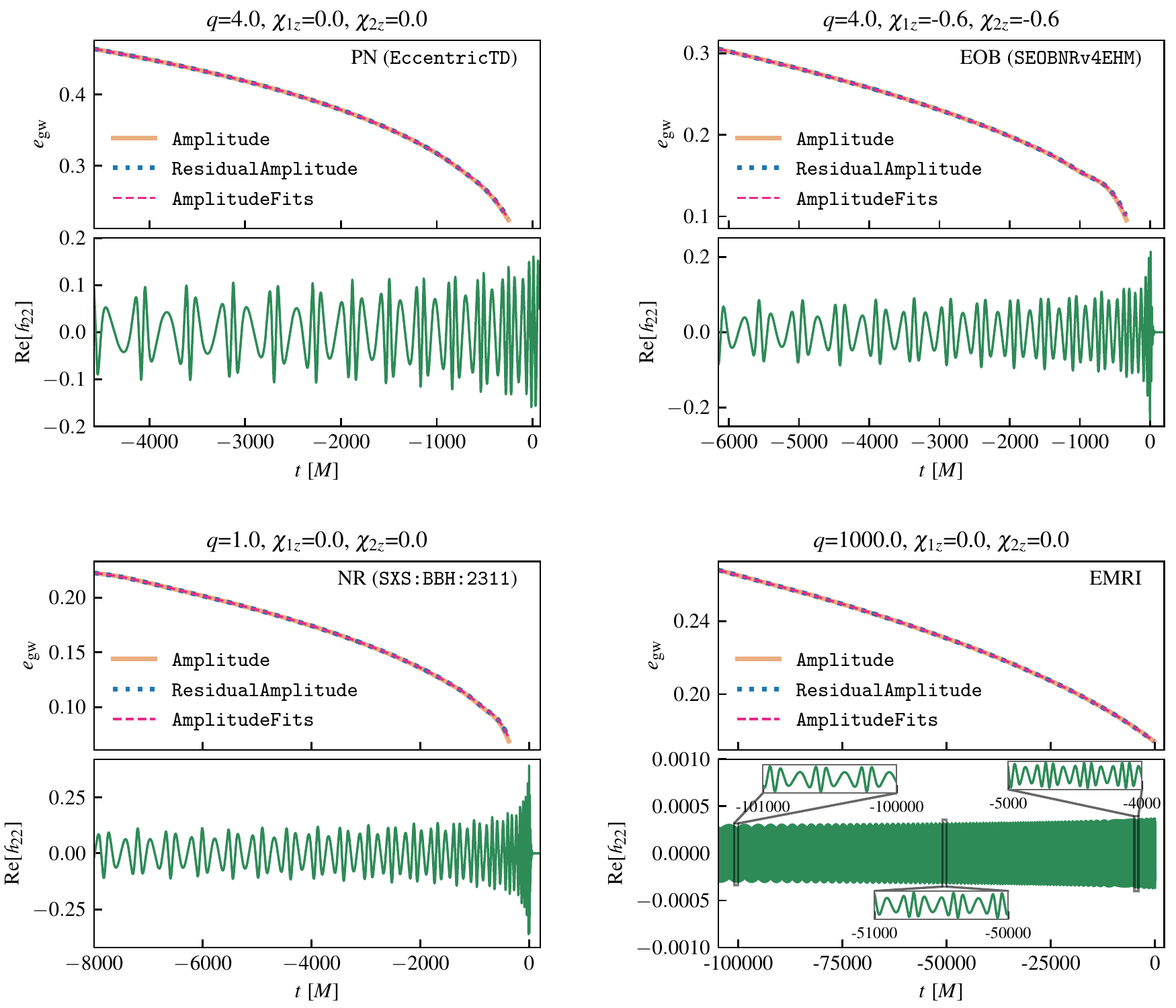}
\caption{
Demonstration of the measurement of eccentricity using the \package{} package
for waveforms of different origins: PN, EOB, NR and EMRI. The binary parameters
are indicated in the figure text. In each subplot, the lower panel shows the
real part of $\htwotwo$, and the upper panel shows the measured eccentricity.
We consider three different methods for identifying the pericenters/apocenters:
\mAmp{}, \mResAmp{} and \mAmpFits{}.
}
\label{fig:demonstrations}
\end{figure*}

\subsection{Applicability for waveforms of different origins}
\label{sec:demonstrations}

Another criteria for the eccentricity definition identified in
Sec.~\ref{sec:introduction} that is desired but not strictly required is
that it should be robust and applicable for waveforms of
different origins, such as analytical PN waveforms~\cite{Memmesheimer:2004cv,
Huerta:2014eca, Tanay:2016zog, Cho:2021oai,Moore:2018kvz, Moore:2019xkm},
numerical waveforms from NR~\cite{Hinder:2017sxy, Boyle:2019kee, Islam:2021mha,
Ramos-Buades:2022lgf, Healy:2022wdn, Habib:2019cui,
Huerta:2019oxn,Hinder:2017sxy, Boyle:2019kee, Ramos-Buades:2022lgf,
Healy:2022wdn, Habib:2019cui, Huerta:2019oxn} simulations, semi-analytical EOB
waveforms calibrated to NR~\cite{Cao:2017ndf, Liu:2019jpg,
Ramos-Buades:2021adz, Nagar:2021gss}, and EMRI~\cite{Warburton:2011fk,
Osburn:2015duj, VanDeMeent:2018cgn, Chua:2020stf, Hughes:2021exa, Katz:2021yft,
Lynch:2021ogr,Barack:2010tm, Akcay:2013wfa, Hopper:2012ty, Osburn:2014hoa,
vandeMeent:2015lxa, Hopper:2015icj, Forseth:2015oua, vandeMeent:2016pee,
vandeMeent:2017bcc, Munna:2020juq, Munna:2020som, Munna:2022gio, Munna:2022xts}
waveforms obtained by solving the Teukolsky equation.

In Fig.~\ref{fig:demonstrations} we show examples of our $\egw$ implementation
in \package{} applied to waveforms of four different origins: PN
(\texttt{EccentricTD}~\cite{Tanay:2016zog}), EOB
(\SEOB{}~\cite{Ramos-Buades:2021adz}), NR
(\SpEC{}~\cite{SXSCatalog,Islam:2021mha}), and EMRI
(Ref.~\cite{Ramos-Buades:2022lgf}). The binary parameters are arbitrarily
chosen to cover a wider parameter space and are shown in the figure text. In
each of the four subplots in Fig.~\ref{fig:demonstrations}, the lower panel
shows the real part of $\htwotwo$, and the upper panel shows the measured
$\egw$. We consider three different methods to locate the
pericenters/apocenters \mAmp{}, \mResAmp{}, and \mAmpFits{}, and $\egw$ is
consistent between the three methods. For the \mResAmp{} method, for the PN,
EOB and EMRI cases, we use the same model evaluated at zero eccentricity for
the quasicircular counterpart. For NR, we use the
\PhenomT{}~\cite{Estelles:2021gvs} model.

In addition to Fig.~\ref{fig:demonstrations}, we have tested our implementation
in \package{} against eccentric \SpEC{} NR waveforms from
Refs.~\cite{Islam:2021mha, Ramos-Buades:2022lgf}. When testing against
eccentric NR simulations from the RIT catalog~\cite{RITCatalog, Healy:2022wdn}, we are able
to compute $\egw$ whenever the waveform contains at least 
$\roughly 4-5$ orbits before the merger, for reasons explained in
Sec.~\ref{sec:defining_eccentricity_using_waveform}. Finally, we have conducted
extensive robustness tests using the \SEOB{} model in different regions of the
parameter space, including converting $\eEOB$ posterior samples to $\egw$
samples in a postprocessing step after parameter estimation.

\subsection{Smoothness tests}
\label{sec:smoothness_tests}

\begin{figure}
\includegraphics[width=\columnwidth]{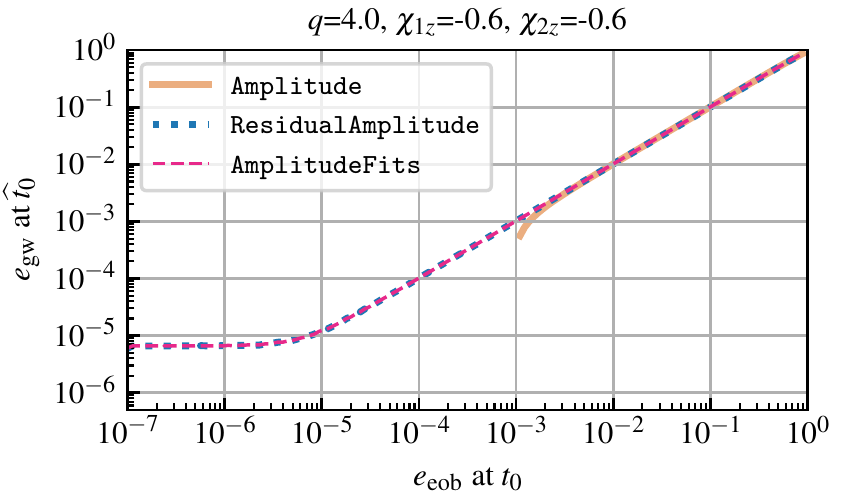}
\caption{
$\egw$ vs $\eEOB$ at the initial time, for \SEOB{} waveforms with varying $\eEOB$,
but keeping the other binary parameters fixed (given in figure title). $\eEOB$
is the model's internal eccentricity, specified at $\tStart=-20000M$. $\egw$ is
evaluated at its first available time, $\tStartHat$. We consider three
different methods for locating pericenters/apocenters: \mAmp{}, \mResAmp{}, and
\mAmpFits{}.  The \mAmp{} method breaks down for small eccentricities ($\eEOB
\lesssim 10^{-3}$), while the \mResAmp{} and \mAmpFits{} method follow
the expected $\egw=\eEOB$ trend down to $\eEOB=10^{-5}$.
}
\label{fig:eob_vs_measured_ecc}
\end{figure}

In this section, we demonstrate that our implementation of $\egw$ varies
smoothly as a function of internal definitions of eccentricity used by waveform
models.  Specifically, we generate 50 waveforms using the \SEOB{}
model~\cite{Ramos-Buades:2021adz}, with the model's internal eccentricity
parameter varying from $\eEOB = 10^{-7}$ to $\eEOB = 0.9$,~\footnote{The upper
limit of $\eEOB=0.9$ is chosen based on the regime of validity of the \SEOB{}
model~\cite{Ramos-Buades:2021adz}, but some tests at higher eccentricity are
included in Sec.~\ref{sec:high_ecc}.} while keeping the other parameters fixed
at $q=4$, and $\chi_{1z}=\chi_{2z}=-0.6$.  The eccentricity $\eEOB$ refers to
the start of each waveform, which we choose to be at $\tStart=-20000M$ before
the peak waveform amplitude.~\footnote{
To achieve the desired length of the inspiral, we adjust the
start frequency of the \SEOB{} model accordingly.
}
In addition to testing whether $\egw$ varies smoothly,
this test also demonstrates that our implementation in \package{} works over a
wide range of eccentricities. Both of these features are important for
applications like converting posterior samples for $\eEOB$ to the standardized
$\egw$.

For simplicity, we restrict our consideration to the three preferred methods
from Sec~\ref{sec:methods_to_locate_extrema}, \mAmp{}, \mResAmp{} and
\mAmpFits{}.  The \mFreq{}, \mResFreq{} and \mFreqFits{} methods perform
similarly to \mAmp{}, \mResAmp{} and \mAmpFits{} methods, respectively, but can
be prone to numerical noise.

\subsubsection{$\egw$ vs $\eEOB$ at initial time}

We first compare $\eEOB$ (which is defined at $\tStart=-20000M$) to $\egw$ at
its first available time (which we denote as $\tStartHat$). As described in
Sec.~\ref{sec:defining_eccentricity_using_waveform}, the first available time
for $\egw(t)$ is the maximum of the times of the first pericenter and first
apocenter, as starting at this time, both $\omegaP(t)$ and $\omegaA(t)$
interpolants in Eq.~\eqref{eq:e_omegatwotwo} can be defined. For our dataset of
\SEOB{} waveforms, this time varies from $\tStartHat=-19250M$ for
$\eEOB=10^{-7}$ to $\tStartHat=-15250M$ for $\eEOB=0.9$. However because the
difference between $\tStartHat$ and $\tStart$ is always within an orbit, and
eccentricity does not change significantly over one orbit, comparing $\egw$ at
$\tStartHat$ to $\eEOB$ at $\tStart$ is reasonable.~\footnote{This assumption
breaks down at very high eccentricity, see Sec.~\ref{sec:high_ecc}.} The ideal
outcome for this test is that the eccentricity measured from the waveform
$\egw$ matches the model's eccentricity definition $\eEOB$.

Figure~\ref{fig:eob_vs_measured_ecc} shows how $\egw$ at $\tStartHat$ varies
with $\eEOB$ at $\tStart$, for the \mAmp{}, \mResAmp{} and \mAmpFits{} methods.
For sufficiently high eccentricities ($\eEOB \gtrsim 5\into10^{-3}$), all three
methods follow the expected trend of $\egw = \eEOB$. However, the \mAmp{}
method starts to deviate from this trend for smaller eccentricities, before
completely breaking down for $\eEOB \lesssim 10^{-3}$. This is expected as
local extrema do not exist in $\Atwotwo$ for such low eccentricities (see
Sec.~\ref{sec:methods_to_locate_extrema}).

\begin{figure}
  \centering
\includegraphics[width=\columnwidth]{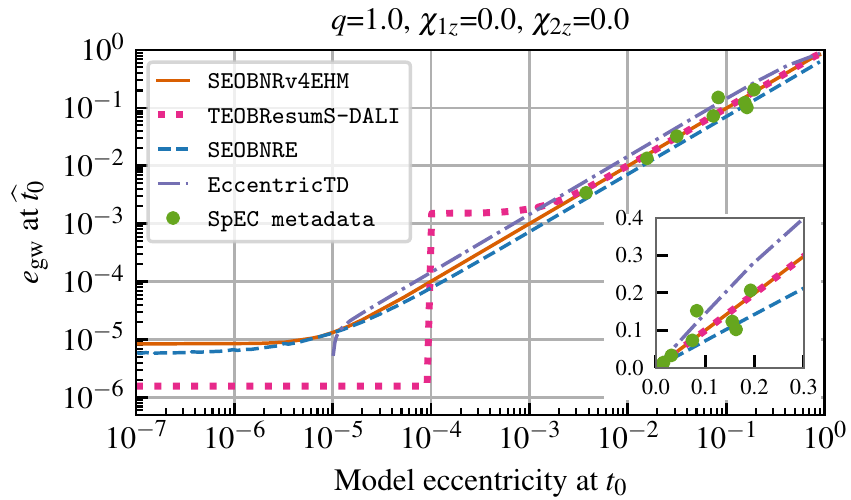}
\caption{
$\egw$ vs the internal definition of eccentricity, for waveforms of different
origin, for equal-mass nonspinning binaries with varying eccentricity.  For the
NR waveforms (SpEC), we compute the internal eccentricity at $\tStart=1500M$
after the start of the simulation, while for the rest we use $\tStart=-20000M$
before peak waveform amplitude. In both cases, $\tStartHat$ is the first
available time for $\egw(t)$. The inset shows the same but on a linear scale,
and focuses on the $\egw\leq0.4$ region.
}
\label{fig:model_ecc_vs_measured_ecc}
\end{figure}

By contrast, the \mResAmp{} and \mAmpFits{} method follow the $\egw=\eEOB$
trend all the way down to $\eEOB = 10^{-5}$. For smaller $\eEOB$, the \SEOB{}
model itself ceases producing waveforms for which the modulations due to
eccentricity decrease with decreasing $\eEOB$. For most practical applications,
this is not problematic for \SEOB{} as $\eEOB=10^{-5}$ is very small. However,
this exercise highlights how (in addition to testing our implementation) tests
like this can help identify the limitations of eccentric waveform models.

In this spirit, we repeat this test for several different eccentric
waveform models in Fig.~\ref{fig:model_ecc_vs_measured_ecc}. For an
equal-mass nonspinning binary, we show how $\egw$ at $\tStartHat$
varies with the internal definitions of eccentricity (defined at
$\tStart=-20000M$) used by the \SEOB{}~\cite{Ramos-Buades:2021adz},
\TEOB{}~\cite{Nagar:2018zoe,Nagar:2021gss},
\SEOBNRE{}~\cite{Cao:2017ndf,Liu:2019jpg}, and
\EccentricTD{}~\cite{Tanay:2016zog} models. For simplicity, we only
consider the \mResAmp{} method, where the quasicircular counterpart is
obtained by evaluating the same model at zero eccentricity.

Figure~\ref{fig:model_ecc_vs_measured_ecc} also shows the dependence
of $\egw$ on the internal definition of eccentricity for a few eccentric
equal-mass nonspinning NR simulations produced with the SpEC
code~\cite{Boyle:2019kee, SXSCatalog,Islam:2021mha} (with SXS IDs 2267, 2270,
2275, 2280, 2285, 2290, 2294 and 2300). In this case, we use the \PhenomT{}
model~\cite{Estelles:2021gvs} for the quasicircular counterpart. The internal
eccentricity for these simulations is computed using the orbital trajectories,
following the method of Refs.~\cite{Buonanno:2010yk, Mroue:2012kv}; we refer to
this as the ``SpEC metadata eccentricity'' as the same method is used to report
eccentricity in the metadata files accompanying the
simulations~\cite{Boyle:2019kee, SXSCatalog}. However, because the publicly
available SpEC metadata files~\cite{SXSCatalog} report eccentricity at
different times for different simulations, we recompute the eccentricity at a
fixed time $\tStart$ using the same methods as Refs~\cite{Buonanno:2010yk,
Mroue:2012kv}. Because the NR simulations are typically short, we choose
$\tStart=1500M$ after the start of the simulations, and $\tStartHat$ (where
$\egw$ is plotted) is once again the first available time for $\egw(t)$.
Before computing $\egw(t)$, the initial parts of the NR waveforms ($t<\tStart$)
are discarded to avoid spurious transients due to imperfect NR initial
data.

In agreement with Fig.~\ref{fig:eob_vs_measured_ecc}, we find that the \SEOB{}
model follows the $\egw=\eEOB$ trend for $\eEOB\gtrsim10^{-5}$ in
Fig.~\ref{fig:model_ecc_vs_measured_ecc}. While \TEOB{} follows the same trend
at higher eccentricities, it deviates significantly from this trend at
$\eEOB\lesssim5\into10^{-3}$, and breaks down at $\eEOB\lesssim10^{-4}$. This
behavior of \TEOB{} was also noted in Ref.~\cite{Knee:2022hth} and suggests
that the model may need improvement in this region. Next, both \SEOBNRE{} and
\EccentricTD{} models fall away from the $y=x$ line in
Fig.~\ref{fig:model_ecc_vs_measured_ecc}, suggesting that the internal
definitions of these models may need modifications. Finally,
the SpEC metadata eccentricity has a scatter around the $y=x$ line. This
behavior is not surprising as the SpEC metadata eccentricity is not meant to be
precise and is known to be sensitive to factors like the length of the time window
used when fitting the orbital trajectories to PN
expressions~\cite{Boyle:2019kee, Buonanno:2010yk, Mroue:2012kv}. Furthermore,
because the orbital trajectories in NR simulations are gauge-dependent, the
eccentricity reported in the SpEC metadata can also be gauge-dependent.  To get
a precise and gauge-independent eccentricity estimate from NR, one must use
waveform-defined quantities like $\egw$.

Figure~\ref{fig:model_ecc_vs_measured_ecc} also shows that for the same $\egw$,
different models have different internal values of eccentricity. Therefore, the
eccentricity inferred from GW signals via Bayesian parameter estimation using
two different models can also be different, highlighting the need for using a
waveform-defined eccentricity like $\egw$. In particular, posterior samples
obtained using different models can be put on the same footing by evaluating
$\egw$ and $\lgw$ using the model's waveform prediction. This approach was
recently taken in Ref.~\cite{Bonino:2022hkj}, albeit restricted to only $\egw$.

%--------------------------------------------------------------------------
\subsubsection{Smoothness of the time evolution of $\egw$}
\label{sec:measured-eccentricity-vs-time}

\begin{figure}
\includegraphics[width=\columnwidth,trim=0 3 0 4]{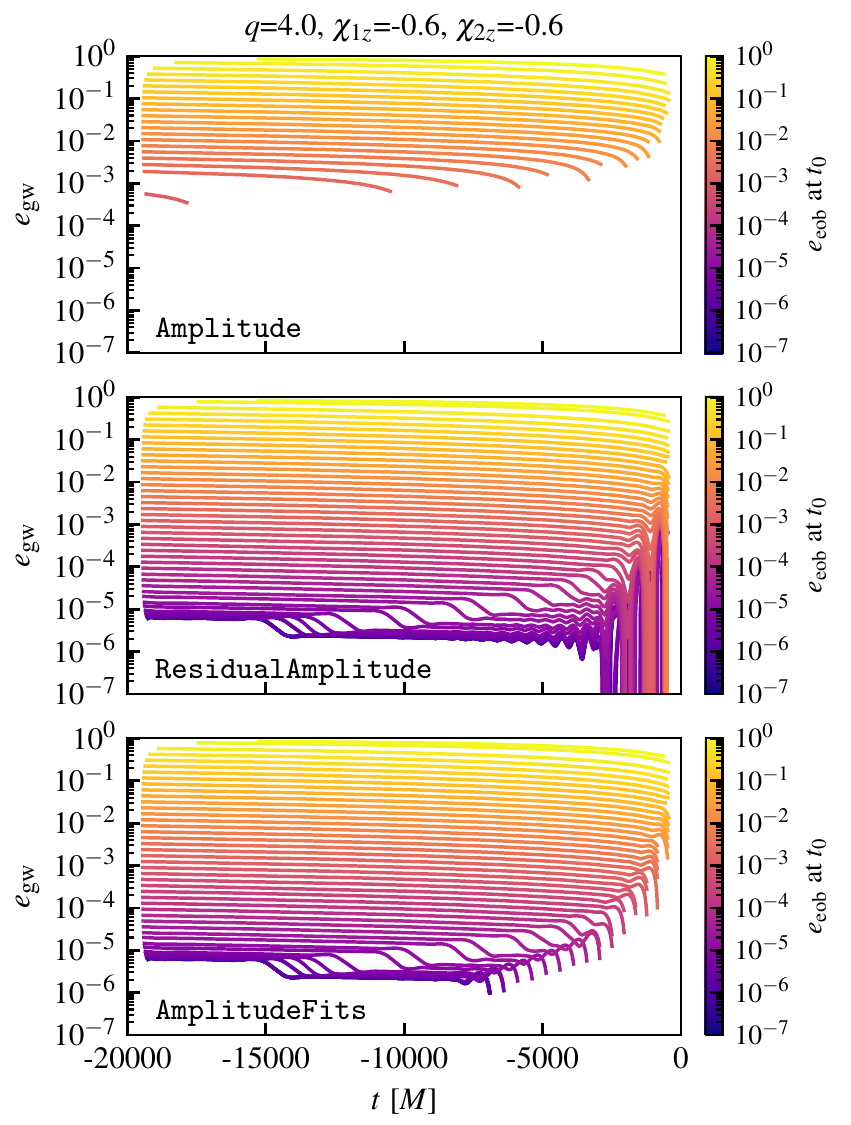}
\caption{
$\egw(t)$ for \SEOB{} waveforms with varying $\eEOB$, but keeping
the other binary parameters fixed (given in figure title). The method used to
locate pericenters/apocenters is indicated in the figure text. The colors
indicate the value of $\eEOB$, defined at $\tStart=-20000M$.  The \mAmp{} method
breaks down for small eccentricities $\egw \lesssim 10^{-3} \ldots 10^{-2}$,
especially as one approaches the merger.  The \mResAmp{} and \mAmpFits{}
methods continue to compute the eccentricity until $\egw \sim10^{-5}$. The
features at $\egw\sim 10^{-5}$ arise from the waveform model itself (see
Fig.~\ref{fig:res_amp_glitches}).
}
\label{fig:measured_ecc_vs_time}
\end{figure}

\begin{figure}
\centering
\includegraphics[width=0.9\columnwidth,trim=0 5 0 6]{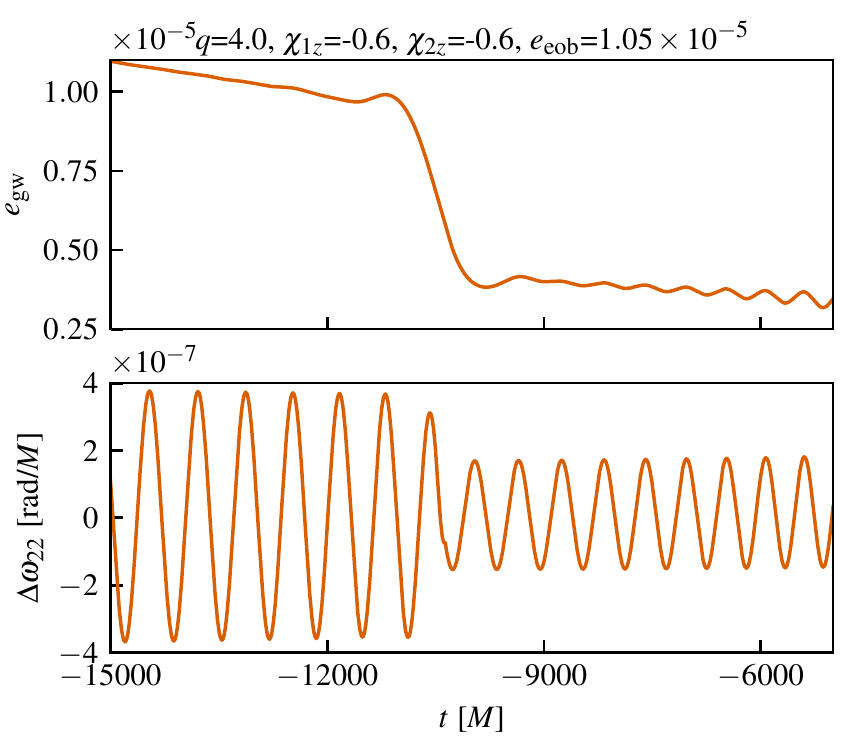}
\caption{
Tracing the noisy features in Fig.~\ref{fig:measured_ecc_vs_time} to the
behavior of the \SEOB{} model at small eccentricities.  The top panel shows
$\egw$ for the case with $\eEOB=1.05 \into 10^{-5}$ at $\tStart = -20000M$,
from the middle panel of Fig.~\ref{fig:measured_ecc_vs_time}.  The bottom panel
shows the corresponding $\resOmega$ (Eq.~\eqref{eq:residual_omega22}), which
helps highlight the modulations due to eccentricity. The drop in $\egw$ occurs
at the same time as an abrupt drop in the eccentricity modulations in
$\resOmega$ that arises from a transition function applied to the
dynamical variables entering the NQC corrections in
\SEOB{}~\cite{Ramos-Buades:2021adz}.
}
\label{fig:res_amp_glitches}
\end{figure}

We now consider a more stringent smoothness test: using the same dataset of 50
\SEOB{} waveforms, we test whether the time evolution of $\egw$ changes
smoothly when varying $\eEOB$ at $\tStart=-20000M$.
Figure~\ref{fig:measured_ecc_vs_time} shows $\egw(t)$ for the \mAmp{},
\mResAmp{} and \mAmpFits{} methods. Even though the waveform data starts at
$\tStart=-20000M$, the $\egw(t)$ is only available for $t\geq\tStartHat$, the
maximum of the times of the first pericenter and apocenter. In
Fig.~\ref{fig:eob_vs_measured_ecc} only eccentricities at the first available
time $\egw(\tStartHat)$ are considered, while in
Fig.~\ref{fig:measured_ecc_vs_time} we consider the full time evolution.

In Fig.~\ref{fig:measured_ecc_vs_time}, we once again find that the \mAmp{}
method breaks down for small eccentricities $\egw \lesssim 10^{-3} \ldots 10^{-2}$,
especially as one approaches the merger as eccentricity is continuously
radiated away. The \mAmp{} method fails when the local extrema in $\Atwotwo$
cease to exist, which is why the curves with smaller initial $\egw$ are
shorter. By contrast, the \mResAmp{} and \mAmpFits{} methods continue to
compute the eccentricity until $\egw \sim10^{-5}$. While the \mResAmp{} method
successfully computes $\egw(t)$ up to the last available orbit (we discard the
last two orbits before the merger as explained in
Sec.~\ref{sec:defining_eccentricity_using_waveform}), the \mAmpFits{} method
misses some extrema near the merger, especially when the eccentricity becomes
small. However, as we will see below, the \mResAmp{} method can depend on the
choice of the quasicircular waveform in the same region.

\begin{figure*}
  \centering
  \includegraphics*[width=\textwidth]{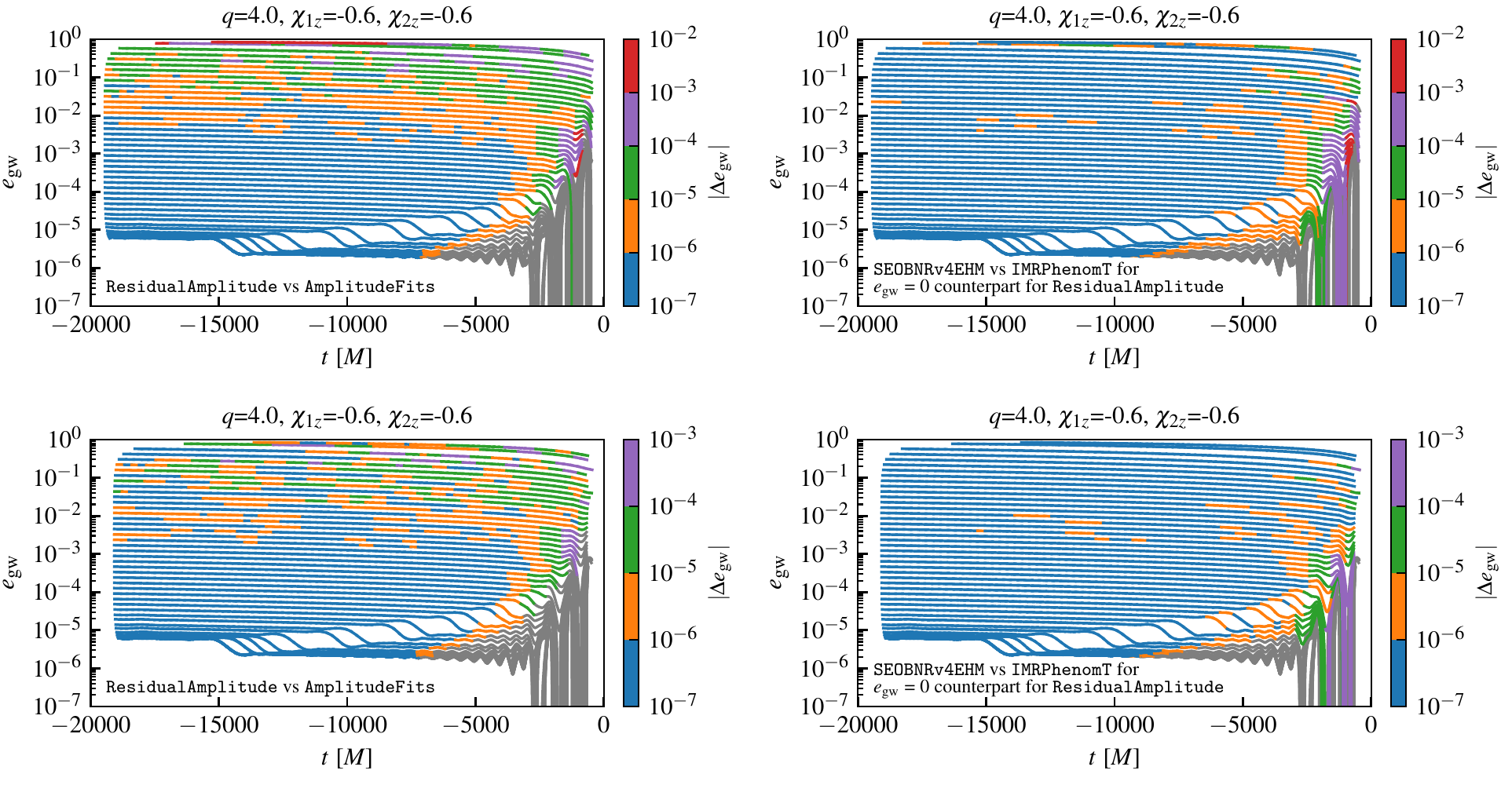}
%   \includegraphics[width=\columnwidth]
%   {compare_measured_eccs_between_methods}
%   \hfill
%   \includegraphics[width=\columnwidth]
%   {compare_measured_eccs_with_diff_zeroecc_waveforms}\\
%   \includegraphics[width=\columnwidth]
%   {compare_measured_eccs_between_methods_with_midpoints}
%   \hfill
%   \includegraphics[width=\columnwidth]
%   {compare_measured_eccs_with_diff_zeroecc_waveforms_with_midpoints}
\caption{
Differences in $\egw(t)$ due to different methods used to locate pericenters and
apocenters, for the same \SEOB{} waveforms as
Fig.~\ref{fig:measured_ecc_vs_time}.
{\itshape Top-left:} The curves show $\egw(t)$ obtained using the \mResAmp{}
method with the quasicircular counterpart also obtained from \SEOB{}. The
colors represent the absolute difference with respect to the $\egw(t)$ obtained
using the \mAmpFits{} method, and the gray region shows the parts where
the second method fails to compute $\egw(t)$.  {\itshape Top-right:} Same, but
now the colors show the difference with respect to the $\egw(t)$ obtained with
\mResAmp{} method with the quasicircular counterpart obtained from the
\PhenomT{} model.  In both top panels, the different choices for locating
pericenters/apocenters lead to broadly consistent results for $\egw(t)$, with
the only notable differences occurring for: (i) small eccentricities ($\egw
\lesssim 5\into10^{-3}$) and near the merger, where the \SEOB{} model also has
known issues (see Fig.~\ref{fig:res_amp_glitches}), and (ii) large
eccentricities ($\egw \sim 0.9$), where locating apocenters is problematic.
The bottom panels show the same as the top panels, but when identifying the
midpoints between pericenters as apocenters. This leads to more consistent
results between different methods, and the largest differences in $\egw$
decrease by an order of magnitude.
}
\label{fig:compare_errors_between_methods}
\end{figure*}

In most regions of Fig.~\ref{fig:measured_ecc_vs_time}, we find that the time
evolution of $\egw$ varies smoothly with $\eEOB$. However, for the \mResAmp{}
and \mAmpFits{} methods, for small eccentricities and near the merger, we find
that $\egw(t)$ can be noisy. Rather than a limitation of these methods, this
behavior arises from the \SEOB{} model itself.
Figure~\ref{fig:res_amp_glitches} focuses on one of the noisy $\egw(t)$ curves
from the middle panel of Fig.~\ref{fig:measured_ecc_vs_time}. The bottom panel
of Fig.~\ref{fig:res_amp_glitches} shows the corresponding $\resOmega(t)$ from
Eq.~\eqref{eq:residual_omega22}, which helps highlight the modulations due to
eccentricity. The fall in $\egw(t)$ is associated with an abrupt fall in the
amplitude of the eccentricity modulations in $\resOmega(t)$.

Such jumps in $\resOmega(t)$ at small eccentricities arise from a transition
function in \SEOB{}~\cite{Ramos-Buades:2021adz} that orbit averages the
dynamical variables entering the non quasicircular (NQC) corrections of the
waveform. The orbit average is carried out between the local maxima of the
oscillations in $\dot{p}_{r^*}$ (see Appendix B of
Ref.~\cite{Ramos-Buades:2021adz} for details). After the last available
maximum, a window is applied to transition from the orbit-averaged variables to
the plunge dynamics (see Eq.~(B2) in Appendix B
of Ref.~\cite{Ramos-Buades:2021adz}). This transition causes the jump in $\Delta
\omegatwotwo$ shown in Fig.~\ref{fig:res_amp_glitches}, as well as the noisy
features at small eccentricity in Fig.~\ref{fig:measured_ecc_vs_time}. Because
the last available maximum occurs at earlier times for smaller eccentricities
---analogous to Eq.~(\ref{eq:e-breakdown})---, these features also start at
earlier times for smaller eccentricities in
Fig.~\ref{fig:measured_ecc_vs_time}. While this behavior is noticeable in our
studies, Ref.~\cite{Ramos-Buades:2023yhy} shows that this causes no
significant biases in parameter estimation, and can be addressed in future
versions of \SEOB{}.  Nevertheless, Fig.~\ref{fig:measured_ecc_vs_time} once
again highlights the importance of such smoothness tests, not only to check our
implementation of $\egw$ but also to identify potential issues in waveform
models.

%--------------------------------------------------------------------------
\subsection{Dependence of $\egw$ on extrema finding methods}
\label{sec:robustness_dependence_on_methods}

For the final robustness test, we consider how strongly $\egw$ depends on the
method used to locate extrema.
We will only consider the \mResAmp{} and \mAmpFits{} methods for simplicity.
From Figs.~\ref{fig:eob_vs_measured_ecc} and \ref{fig:measured_ecc_vs_time}, we
already see that $\egw$ is broadly consistent between different methods. We now
quantify the differences in Fig.~\ref{fig:compare_errors_between_methods}, for
the same dataset of 50 \SEOB{} waveforms from Sec.~\ref{sec:smoothness_tests}.

The top-left panel of Fig.~\ref{fig:compare_errors_between_methods} shows
$\egw(t)$ for these waveforms when using the \mResAmp{} method and
the colors represent the instantaneous absolute difference
with respect to the $\egw(t)$ obtained from the \mAmpFits{} method.
Here, we use \SEOB{} evaluated at zero eccentricity for the
quasicircular counterpart required for \mResAmp{}. The gray region
represents the parts where \mResAmp{} can compute $\egw(t)$,
but \mAmpFits{} can not. However, we note that this only occurs for
small eccentricities $\egw \lesssim 5\into10^{-3}$, and close to
the merger. This region also coincides with the region where \SEOB{}
exhibits the noisy behavior discussed in
Fig.~\ref{fig:res_amp_glitches}.

Next, the top-right panel of
Fig.~\ref{fig:compare_errors_between_methods} illustrates the
difference in $\egw(t)$ between different choices of quasicircular
counterpart for the \mResAmp{} method. The curves once again represent
$\egw(t)$ evaluated using \mResAmp{} with the quasicircular
counterpart obtained from \SEOB{} (the same model used to produce the
eccentric waveforms). The colors represent the instantaneous
absolute difference with respect to the $\egw(t)$ obtained from the \mResAmp{}
method with the quasicircular counterpart obtained from the \PhenomT{}
model instead. The gray region represents the parts
where \mResAmp{} using \SEOB{} for the quasicircular counterpart can compute $\egw(t)$, but \mResAmp{} using \PhenomT{} can
not. Once again, this occurs only for small eccentricities and near
the merger.  In this regime, the small differences between \SEOB{} (in
the quasicircular limit) and \PhenomT{}, especially near the merger,
become important, and \PhenomT{} does not accurately capture the
secular growth in \SEOB{}.

In the regions where both \mResAmp{} and \mAmpFits{} methods successfully
compute $\egw(t)$ in the top-left panel of
Fig.~\ref{fig:compare_errors_between_methods}, the biggest differences are of
order $10^{-2}$. These differences occur either for small eccentricities near
the merger, or for very large eccentricities ($\egw \sim 0.9$). At such high
eccentricities, the waveform is characterized by sharp bursts at pericenter
passages alternating with wide valleys that include the apocenter passages (see
bottom panel of Fig.~\ref{fig:omega22_average}, for example). As a result, it
is easy to identify the pericenter times but not the apocenter times for these
waveforms. This can be resolved by only identifying the pericenter times and
\emph{defining} the apocenter times to be the midpoints between consecutive
pericenters. The assumption employed here is that the radiation reaction is not
strong enough that the times taken for the first and second halves of an orbit
are significantly different.  While this assumption is broken near the merger,
we already discard the last two orbits before the merger when computing $\egw$
(Sec.~\ref{sec:defining_eccentricity_using_waveform}). 

The bottom panels of Fig.~\ref{fig:compare_errors_between_methods} show the
same as the top panels, but when identifying the midpoints between pericenters
as apocenters. We find that the largest differences between \mResAmp{} and
\mAmpFits{}, as well as the largest differences between \mResAmp{} with
different quasicircular counterparts, are now an order of magnitude smaller.
This suggests that identifying the midpoints between pericenters as apocenters
may be a more robust choice than directly locating apocenters, especially for
large eccentricities. We provide this as an option in \package{}. 

To summarize, the different choices for locating extrema in
Fig.~\ref{fig:compare_errors_between_methods} lead to broadly consistent
results for $\egw(t)$, with the only notable differences occurring for: (i)
small eccentricities ($\egw \lesssim 5\into10^{-3}$) and near the merger, where
the \SEOB{} model also has known issues (see Fig.~\ref{fig:res_amp_glitches}),
and (ii) large eccentricities ($\egw \sim 0.9$), where locating apocenters is
problematic. As discussed in Sec.~\ref{sec:methods_to_locate_extrema}, such
differences are expected, and the different methods to locate extrema should be
regarded as different definitions of eccentricity. However, identifying the
midpoints between pericenters as apocenters, rather than directly locating
apocenters, can lead to more consistent results between different methods.

\subsection{Applicability for the high eccentricity regime}
\label{sec:high_ecc}

\begin{figure}
  \centering
  \includegraphics{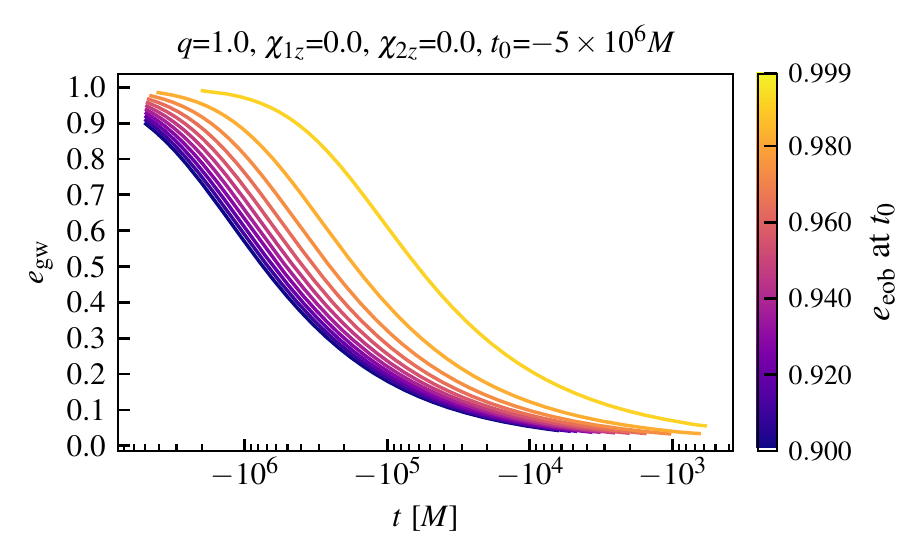}
\caption{
A smoothness test for $\egw(t)$ at very high eccentricities. The curves show
the time evolution of $\egw$ for \SEOB{} waveforms with initial eccentricities
$0.9 \leq \eEOB \leq 0.999$ at $\tStart=-5\into10^{6}M$. The colors
represent $\eEOB$ at $\tStart$. The binary parameters are shown in the figure
title. We use the \mResAmp{} method to locate pericenters and identify the
midpoints between pericenters as apocenters.
}
\label{fig:high_ecc}
\end{figure}
The tests we have conducted so far have been restricted to $\egw\leq0.9$. In
this section, we focus on testing our implementation in the high eccentricity
regime, $\egw > 0.9$. While the eccentricity definition adopted in this work
is, in principle, valid at all eccentricities in the range $(0 - 1)$, high
eccentricity comes with additional challenges:
\begin{itemize}
    \item As $\egw\to1$, the separation in time between pericenters increases
        (see Fig.~\ref{fig:omega22_average}), making it challenging to produce
        waveforms with enough extrema to construct the $\omegaP(t)$ and
        $\omegaA(t)$ interpolants required in Eq.~(\ref{eq:e_omegatwotwo}).
        This limits the practical applicability of $\egw$ for high-eccentricity
        NR simulations like those in Ref.~\cite{Healy:2022wdn}.  However, as we
        will see below, if sufficiently long waveforms can be produced, this
        definition of eccentricity and our implementation still work at high
        eccentricities.
    \item The first available time $\tStartHat$ for $\egw(t)$, is the maximum
        of the times of the first pericenter and first apocenter
        (Sec.~\ref{sec:defining_eccentricity_using_waveform}), which occurs up
        to an orbit after the start of the waveform, $\tStart$. As the duration
        of orbits increases with eccentricity, so does the difference between
        $\tStart$ and $\tStartHat$. As eccentricity also evolves during this
        time, a non-negligible amount of eccentricity may be radiated away
        before the first available time for the $\egw(t)$ measurement (see
        an example below).
    \item As discussed in Sec.~\ref{sec:robustness_dependence_on_methods},
        locating apocenters becomes challenging at high eccentricities.
        Therefore, in the test below, we identify the midpoints between
        pericenters as apocenters, rather than directly locating apocenters.
\end{itemize}

To test our implementation at high eccentricities, we construct a new dataset
of \SEOB{} waveforms with eccentricities $0.9 \leq \eEOB \leq 0.999$ defined at
$\tStart=-5\into10^{6}M$,~\footnote{Once again, we achieve the desired length of
the inspiral by adjusting the start frequency of the \SEOB{} model
accordingly.} for a system with parameters $q=1$, and $\chi_{1z}=\chi_{2z}=0$.
For this dataset, the waveforms include 154 (672) pericenters before
merger for $\eEOB=0.999$ ($\eEOB=0.9$) at $\tStart$, allowing us to easily
measure $\egw(t)$ even for such high eccentricities.  Figure~\ref{fig:high_ecc}
shows the eccentricities $\egw(t)$ measured using these waveforms; as expected,
$\egw(t)$ varies smoothly with varying $\eEOB$ even for high eccentricities.
For the waveform with $\eEOB=0.999$ at $\tStart$, the measured eccentricity at
the first available time $\tStartHat\approx -2\into 10^{6}M$ is $\egw \approx
0.99$. Because the gap between $\tStartHat$ and $\tStart$ is very large
($\roughly3\into 10^{6}M$) for this case, it is unsurprising that the
eccentricity decays to $\egw(\tStartHat)=0.99$. The choice of
$\tStart=-5\into10^{6}M$ was made for this dataset so that $\tStart$ occurs
early in the inspiral, where this decay is less drastic. 

%==========================================================================
\section{Conclusion}
\label{sec:conclusion}
We present a robust implementation of standardized definitions of
eccentricity ($\egw$) and mean anomaly ($\lgw$) that are computed directly from
the gravitational waveform (Sec.~\ref{sec:defining_eccentricity}). Our method
is free of gauge ambiguities, has the correct Newtonian limit, and is
applicable for waveforms of all origins, over the full range of allowed
eccentricities for bound orbits. However, as our method relies on computing the
frequency at pericenter and apocenter passages, it requires waveforms with at
least $\roughly 4-5$ orbits.

Our method can be applied directly during source parameter estimation or as a
postprocessing step to convert posterior samples from the internal definitions
used by models and simulations to the standardized ones. This puts all models
and simulations on the same footing, while also helping connect GW observations
to astrophysical predictions for GW populations. Finally, we propose how the
reference frequency $\fref$ and start frequency $\fLow$, that are used in GW
data analysis, should be generalized for eccentric binaries
(Secs.~\ref{sec:generalizing_fref}, \ref{sec:selecting_reference_point},
\ref{sec:generalizing_fLow}).

One key aspect of computing $\egw$ and $\lgw$ is identifying the times of
pericenter and apocenter passages from the waveform. We provide different
methods for this purpose, that should be treated as different variants of the
eccentricity definition. Among the provided methods (see
Sec.~\ref{sec:methods_to_locate_extrema}), the \mAmp{} method is applicable
when eccentricity is sufficiently high ($\egw \gtrsim 10^{-3} \ldots 10^{-2}$),
while \mResAmp{} and \mAmpFits{} are applicable for smaller eccentricities as
well.

We demonstrate the robustness of our implementation by testing against
waveforms of different origins, including PN, EOB, EMRIs and NR
(Sec.~\ref{sec:demonstrations}).  We further conduct smoothness tests that have
the added benefit of identifying noisy features in waveform models
(Sec.~\ref{sec:smoothness_tests}). Our tests include waveforms with
eccentricities ranging from $10^{-5}$ to 0.999. We discuss the limitations
of our approach for very high eccentricities (Sec.~\ref{sec:high_ecc}),
especially for NR simulations where including $\roughly 4-5$ orbits can be
challenging when eccentricity is high.

We make our implementation publicly available through an easy-to-use \python{}
package, \package{}. This work focuses on systems without spin-precession, and
the most important next step is to generalize our methods to spin-precessing
eccentric binaries. We leave this to future work but discuss potential
approaches (Sec.~\ref{sec:extending_to_precessing_and_FD}).

%==========================================================================
\begin{acknowledgments}
% Randos
We thank Peter James Nee and Leo C. Stein for useful discussions and Geraint
Pratten, Isobel Romero-Shaw, Teagan Clarke, Paul Lasky, Eric
Thrane and Aditya Vijaykumar for comments on the manuscript.
% M.A.S
M.A.S.’s research was supported by the Department of Atomic Energy,
Government of India and the National Research Foundation of Korea
under grant No.~NRF-2021R1A2C2012473. M.A.S acknowledges travel
support from the Infosys Exchange Scholars program to visit AEI, Potsdam
and hospitality by AEI, Potsdam where a part of the work was
completed.
% V.V
V.V acknowledges support from the European Union’s Horizon 2020
research and innovation program under the Marie Skłodowska-Curie grant
agreement No.~896869.
% M.v.d.M
M.v.d.M. is supported by VILLUM FONDEN (grant no. 37766), and the Danish Research Foundation.
% LIGO
This material is based upon work supported by NSF's LIGO Laboratory
which is a major facility fully funded by the NSF.
% GWOSC
%This research made use of data, software and/or web tools obtained from the
%Gravitational Wave Open Science Center~\cite{GW_open_science_center}, a service
%of the LIGO Laboratory, the LIGO Scientific Collaboration and the Virgo
%Collaboration.
% Computing resource
Most of the numerical calculations reported in this paper as well as
the development of \package{} were performed using the
Alice cluster at ICTS-TIFR.
\end{acknowledgments}

%%%%%%%%%%%%%%%%%%%%%%%%%%%%%%%%%%%%%%%%%%%%%%%%%%%%%%%%%%%%%%%%%%%%%%%%%%%%%%%
\section*{References}
%%%%%%%%%%%%%%%%%%%%%%%%%%%%%%%%%%%%%%%%%%%%%%%%%%%%%%%%%%%%%%%%%%%%%%%%%%%%%%%
\bibliography{References}

\end{document}